\documentstyle[prb,aps]{revtex}
\begin{document}

\title{Single-Particle Green Functions in Exactly Solvable Models 
 of Bose and Fermi Liquids}
\author{Girish S. Setlur and Yia-Chung Chang}
\address{Department of Physics and Materials Research Laboratory,\\
 University of Illinois at Urbana-Champaign , Urbana Il 61801}
\maketitle

\begin{abstract}
 Based on a class of exactly solvable models of interacting bose and fermi
 liquids, we compute the single-particle propagators of these systems
 exactly for all wavelengths and energies and in any number of
 spatial dimensions. The field operators are expressed
 in terms of bose fields that correspond to displacements of the condensate
 in the bose case and displacements of the fermi sea in the fermi case.
 Unlike some of the previous attempts, the present attempt reduces
 the answer for the spectral function in any dimension 
 in both fermi and bose systems to quadratures.
 It is shown that when only the lowest order sea-displacement
 terms are included, the random phase approximation in its many guises
 is recovered in the fermi
 case, and Bogoliubov's theory in the bose case. The momentum distribution 
 is evaluated using two different approaches, 
 exact diagonalisation
 and the equation of motion approach.
 The novelty being of course,
 the exact computation of single-particle properties including
 short wavelength behaviour.                              
\end{abstract}

\section{Introduction}
 Recent years have seen remarkable developments in many-body theory
 in the form of an assortment of techniques that may be loosely termed
 bosonization. The beginnings of these types of techniques may be
 traced back to the work of Tomonaga\cite{Tom}
 and later on by Luttinger \cite{Lutt} and by Lieb and Mattis
 \cite{Lieb}. The work of Sawada \cite{Sawada} and Arponen and Pajanne
 \cite{Arponen} in recasting the fermi gas problem in a bose language has
 to be mentioned.  Arponen and Pajanne recover corrections to
 the Random Phase Approximation (RPA) of Bohm and Pines
 \cite{Bohm} in a systematic manner. In nuclear physics, bosonization
 is widely used to study collective properties, for an introduction
 see the book by Iachello and Arima \cite{Arima}.
 In the 70's an attempt was made by Luther \cite{Luther}
 at generalising these ideas to higher dimensions. Closely related to this
 is the work by Sharp et. al. \cite{Sharp} in current algebra.
 More progress was made by Haldane \cite{Haldane} which culminated in
 the explicit computation of the single-particle propagator by
 Castro-Neto and Fradkin \cite{Neto} and by Houghton, Marston et.al.
 \cite{Mars} and also by Kopietz et. al. \cite{Kop}.
 Rigorous work by Frohlich and Marchetti\cite{Froch}
 is also along similar lines. Also the work of Frau et. al. \cite{Zemba}
 on algebraic bosonization is relevant to the present article as the
 authors have considered effects beyond the linear dispersion in
 that article. The exactly solvable models of Calogero and Sutherland
 are of relevance here as well,
 the exact propagators of these models have been computed
 by various authors \cite{Lesage}. Recently, these types of models have
 been generalised to more than one dimension by Ghosh \cite{Ghosh}.

 The attempt made here is to generalise the concepts of Haldane \cite{Haldane}
 to accomodate short-wavelength fluctuations where the concept of
 a linearised bare fermion energy dispersion is no longer valid.
 To motivate progress in this direction, we find that it is necessary to
 introduce two different concepts, one is the canonical
 conjugate of the fermi/bose density distribution, the other is the concept
 of sea/condensate displacements. 

 Histrorically speaking, the idea that the
 velocity operator could serve as the canonical
 conjugate of the density has been around for a long time, and this
 has been exploited in the study of HeII by Sunakawa et. al. \cite{Sun}.
 However, the authors are not aware of a rigorous study of the meaning
 of this object, in particular, an explicit formula for the
 canonical conjugate of the density operator has to the best of the 
 authors' knowledge never been written down in terms of the field operators.
 The work by Sharp et. al. \cite{Sharp} comes close to what we are
 attempting here.

 The concept of a sea-displacement is a
 generalisation of the traditional approach used for bosonizing one-dimensional
 systems such as the Tomonaga-Luttinger{\cite{Tom},\cite{Lutt}} models.
 There, one introduces bose
 fields that correspond physically to displacement of the fermi surface
 (in 1D, fermi points). These bose fields have simple forms relating them
 to number-conserving product of fermi fields. The field operator is obtained
 by exponentiating the commutation rule between the surface-displacement
 operator
 and the field operator. By analogy, we generalise these ideas, so that one
 is no longer restricted to be close to the fermi surface. The way this is done
 is to postulate the existence of bose fields that correspond to displacements
 of the fermi-sea rather than just the fermi surface. From this it is possible
 to write down formulas for the number-conserving product of two fermi-fields 
 in terms of the bose fields. A similar construction is possible when the
 parent fields are bosons, but here, we find that instead of
 sea-displacements, we have to introduce operators that correspond
 physically to displacements of the condensate. Actually, the bose case is
 much simpler and a mathematically rigorous formulation of this
 correspondence is possible. This is a boon, since we use this fact and 
 make plausible the analogous correspondence in the fermi case.
 The assertions in the fermi case are not proved
 "rigorously", rather are made exceedingly plausible by analogy.
 This is the main drawback of this article.
 
 This article is organised as follows. In the next section, we present
 some formulas that relate the number conserving product of two 
 fermi/bose fields to the relevant sea/condensate-displacement operators
 that are postulated to be canonical bosons. The sea/condensate-displacement
 operators in turn may be related to the parent fermi/bose fields, as
 it happens, this formula is simple in the case when the parent fields are
 bosons but is difficult in the case when the parent fields are fermions.
 
 Following this, we write down
 a generic formula for the fermi/bose field-operator in terms of the density
 operator (operator-valued distribution, to be precise) and its canonical
 conjugate. The new ingredient in this section is the canonical conjugate
 on the density operator. This quantity may in turn be related to currents
 and densities. We find that these formulas are ambiguous unless a proper
 choice is made for a certain phase functional. For bosons, we find that
 this choice is the zero-functional but for fermions it has to be determined
 by making contact with the free theory(done in the section following the
 one just described).

 Combining the two previous sections, we
 write down in the next section, formulas for currents
 and densities in terms of the sea/condensate-displacements, 
 the field-operator has a formula in terms of the
 sea/condensate-displacements as well. Contact is made with the
 propagator of the free theory and the undetermined phase functional
 of the previous sections
 is determined for the fermi case. In the next section, interaction terms are
 introduced that correspond to two-body repulsive interactions.
 It is argued
 and demonstrated that selectively retaining parts of the interaction 
 that are quadratic in the sea/condensate-displacements 
 amounts to doing Bogoliubov/RPA theory. Corrections to this quadratic 
 hamiltonian are easy to write down but are not used to compute corrections
 to RPA/Bogoliubov theory as this requires significantly more effort.
 It is found that the diagonalisation of the RPA hamiltonian
 is rather tricky if one wants to recover both the particle-hole modes
 and the collective mode. In the end, closed formulas are written down for
 the fermi propagator in all three spatial dimensions and their various
 qualitative features are examined. 
 This completes the solution of the
 many-body problem in the RPA/Bogoliubov-limit.

 The appendices are as follows, Appendix A contains a detailed proof of the
 correspondence between the number-conserving product of two bose fields
 and the corresponding condensate-displacements. Appendix B involves 
 writing down similar ideas for fermi systems. However here, the various
 assertions are only made plausible unlike in the bose case where a 
 rigorous solution is possible.
 Appendix C is devoted to 
 proving the assertion that retaining only terms linear in the
 sea-displacements in the definition of the density recovers the RPA. 
 Appendix D involves a derivation of the formula for the momentum distribution
 of the 1D system using the equation of motion approach. Appendix E
 contains some technical statements regarding the proof of uniqueness
 of the formula relating the fermi field with the corresponding currents
 and densities.

\section{Expressing Products of Parent Fields in Terms of Sea-displacements}
 In this section we introduce canonical bose fields called sea-displacements
 in the fermi case and condensate-displacements in the bose case. 
 First, we write down a formula for the number conserving product of
 two bose fields in terms of the condensate displacement operators. 
 A rigourous proof of this is relegated to Appendix A.
 The correspondence is made plausible by making several observations
 about these formulae. Let us first focus on the bose case. Let
 $ b_{ {\bf{q}} } $ and $ b^{\dagger}_{ {\bf{q}} } $ be canonical 
 bose operators. From these, we may construct other bose operators defined
 as follows($ {\bf{q}} \neq {\bf{0}} $),
\begin{equation}
d_{ {\bf{q}}/2 }({\bf{q}}) = (\frac{1}{\sqrt{N_{0}}})
b^{\dagger}_{ {\bf{0}} }b_{ {\bf{q}} }
\end{equation}
and,
\begin{equation}
d_{ {\bf{0}} }({\bf{0}}) = 0
\end{equation}
 where $ N_{0} = b^{\dagger}_{ {\bf{0}} }b_{ {\bf{0}} } $.
 This is the condensate displacement annhilation operator. It is so named
 for the following reason. The definition suggests that this operator
 removes a boson from among those that are not in the condensate 
 and returns it to the condensate, thereby displacing the latter.  
 The reason for the redundant momentum label in the notation
 $ d_{ {\bf{q}}/2 }({\bf{q}}) $ becomes clear if one realises
 that a more general object would be a sea-displacement annhilation
 operator $ d_{ {\bf{k}} + {\bf{q}}/2 }({\bf{q}}) $ . Since
 the condensate corresponds to $ {\bf{k}} = 0 $, we have just the
 condensate displacement annhilation operator. In fact, it will be shown
 subsequently that for the fermi case we have to deal with this
 more general object namely, the sea-displacement annhilation operator.
 It may be shown that (see Appendix A) this object
 $ d_{ {\bf{q}}/2 }({\bf{q}}) $ satisfies canonical bose commutation rules. 
 Also a formula is possible for the number-conserving
 product of two parent bosons in terms of these condensate-displacements.
 The formula is written down below and proved in Appendix A.
\[
b^{\dagger}_{ {\bf{k+q/2}} }b_{ {\bf{k-q/2}} }
 =
N_{0}
 \delta_{ {\bf{k}}, 0 }\delta_{ {\bf{q}}, 0 } +
[\delta_{ {\bf{k+q/2}}, 0 }(\sqrt{N_{0}})d_{ {\bf{k}} }(-{\bf{q}})
 + \delta_{ {\bf{k-q/2}}, 0 }d^{\dagger}_{ {\bf{k}} }({\bf{q}})
(\sqrt{N_{0}})]
\]
\begin{equation}
+
d^{\dagger}_{ (1/2)({\bf{k+q/2}}) }({\bf{k+q/2}})
d_{ (1/2)({\bf{k-q/2}}) }({\bf{k-q/2}})
\label{BOSE}
\end{equation}
where,
\begin{equation}
N_{0} = N - \sum_{ {\bf{q}}_{1} }d^{\dagger}_{ {\bf{q}}_{1}/2 }({\bf{q}}_{1})
d_{ {\bf{q}}_{1}/2 }({\bf{q}}_{1})
\end{equation}
and,
\begin{equation}
[d_{ {\bf{q}}/2 } ({\bf{q}}), N] = 0
\end{equation}
\begin{equation}
N = \sum_{ {\bf{q}} }b^{\dagger}_{ {\bf{q}} }b_{ {\bf{q}} }
\end{equation}
also the object $ d_{ {\bf{0}} }({\bf{0}}) = 0 $, by definition.

 The way the authors initially derived this formula is as follows.
 One starts off with the observation that the object 
 $ b^{\dagger}_{ {\bf{k}} + {\bf{q}}/2 } b_{ {\bf{k}} - {\bf{q}}/2 } $ 
 is the only one that enters in the hamiltonian of number-conserving
 systems. Furthermore,  it satisfies closed commutation rules amongst
 other members of its kind. One is therefore led to look for formulas
 for these objects in terms of other bosons with a view to make the 
 full hamiltonian more easily diagonalisable. In particular, if there
 were bose operators $ d_{ {\bf{q}}/2 }({\bf{q}}) $ and 
 $ d^{\dagger}_{ {\bf{q}}/2 }({\bf{q}}) $ such that
 $ b^{\dagger}_{ {\bf{k}} + {\bf{q}}/2 } b_{ {\bf{k}} - {\bf{q}}/2 } $ was
 exactly linear in these bosons, then the full hamiltonian would indeed
 be exactly diagonalisable. We find that this is not the case and there are
 corrections to this linear term and it so happens that introduction of a 
 quadratic term in the condensate-displacements in fact makes the 
 correspondence exact. The authors are not aware of a deeper reason behind
 this simple formula that terminates after including the quadratic term,
 after all, the formula for the parent annhilation operator 
 $ b_{ {\bf{q}} } $ in terms of the condensate displacements is
 formidable as we shall soon see. The bose case being so simple and
 exact can be used as a benchmark to write down corresponding formulas in the
 fermi case, where rigorous proofs are much harder to come by. The authors
 also
 have in mind generalisations to relativistic systems, where one might
 profit by following the above prescription. In particular, it would
 be fascinating to see if the ideas above were useful in getting 
 non-perturbative information regarding gauge theories like QED, QCD e.t.c.
 But this is far into the future. For now, let us try to write down
 a similar correspondence for the nonrelativistic fermi system.

 As mentioned earlier, for fermi systems, it is necessary to postulate the
 existence of a sea-displacement annhilation operator, denoted
 by $ a_{ {\bf{k}} }({\bf{q}}) $. A formula for this in terms of the
 fermi fields is extremely difficult to deduce. In Appendix B,
 attempts are made to do exactly this. There it is pointed out that these
 objects satisfy canonical boson commutation rules.
  The important issues that
 enable one to draw practical conclusions, fortunately do not depend
 very much on the technical details. In Appendix B and in the
 sections that follow, we show how to extract
 the necessary physics while cirumventing the technical details.
 It must be pointed out however that this drawback is regrettable.
 Let us merely quote the final answers and later on make these formulas
 plausible.
 The RPA form of the number conserving product of two fermi fields
 in terms of the sea-bosons is given by(  $ {\bf{q}} \neq 0 $ ),
\[
c^{\dagger}_{ {\bf{k}} + {\bf{q}}/2 }
c_{ {\bf{k}} - {\bf{q}}/2 }
 = 
(\frac{N}{\langle N \rangle })^{\frac{1}{2}}
[\Lambda_{ {\bf{k}} }({\bf{q}})
a_{ {\bf{k}} }(-{\bf{q}})
+ a^{\dagger}_{ {\bf{k}} }({\bf{q}})
\Lambda_{ {\bf{k}} }(-{\bf{q}})]
\]
\[
+ T_{1}({\bf{k}}, {\bf{q}})\sum_{ {\bf{q}}_{1} }
a^{\dagger}_{ {\bf{k}} + {\bf{q}}/2 - {\bf{q}}_{1}/2 }({\bf{q}}_{1})
a_{ {\bf{k}} - {\bf{q}}_{1}/2 }({\bf{q}}_{1} - {\bf{q}})
\]
\begin{equation}
- T_{2}({\bf{k}}, {\bf{q}})\sum_{ {\bf{q}}_{1} }
a^{\dagger}_{ {\bf{k}} - {\bf{q}}/2 + {\bf{q}}_{1}/2 }({\bf{q}}_{1})
a_{ {\bf{k}} + {\bf{q}}_{1}/2 }({\bf{q}}_{1} - {\bf{q}})
\end{equation}
here,
\begin{equation}
T_{1}({\bf{k}}, {\bf{q}}) = \sqrt{ 1 - {\bar{n}}_{ {\bf{k}} + {\bf{q}}/2 } }
\sqrt{ 1 - {\bar{n}}_{ {\bf{k}} - {\bf{q}}/2 } }
\end{equation}
\begin{equation}
T_{2}({\bf{k}}, {\bf{q}}) = \sqrt{  {\bar{n}}_{ {\bf{k}} + {\bf{q}}/2 } 
{\bar{n}}_{ {\bf{k}} - {\bf{q}}/2 } }
\end{equation}
\begin{equation}
\Lambda_{ {\bf{k}} }({\bf{q}}) = \sqrt{ {\bar{n}}_{ {\bf{k}} + {\bf{q}}/2 }
(1 - {\bar{n}}_{ {\bf{k}} - {\bf{q}}/2 }) }
\end{equation}
here, the sea-boson commutes with the total number of fermions,
\begin{equation}
[a_{ {\bf{k}} }({\bf{q}}), N] = 0
\end{equation}
and also the operator $ a_{ {\bf{k}} }({\bf{0}}) = 0 $.
Further,
\begin{equation}
n_{ {\bf{k}} } = n^{\beta}({\bf{k}}) \frac{ N }{ \langle N \rangle }
+ \sum_{ {\bf{q}} }a^{\dagger}_{ {\bf{k}} - {\bf{q}}/2 }({\bf{q}})
a_{ {\bf{k}} - {\bf{q}}/2 }({\bf{q}})
 - \sum_{ {\bf{q}} }a^{\dagger}_{ {\bf{k}} + {\bf{q}}/2 }({\bf{q}})
a_{ {\bf{k}} + {\bf{q}}/2 }({\bf{q}})
\label{NUMBER}
\end{equation}
and 
\begin{equation}
n^{\beta}({\bf{k}}) = \frac{1}{ exp(\beta(\epsilon_{ {\bf{k}} }-\mu)) + 1 }
\end{equation}
 Also $ {\bar{n}}_{ {\bf{k}} } = \langle n_{ {\bf{k}} } \rangle $.
 The expectation value is with respect to the full interacting ground state.
 This quantity depends on the interactions that are
 present in the system and
 must be evaluated self-consistently. In fact, there is a deeper reason
 for introducing this. The exact formula for the number conserving product
 of two fermi fields and the sea-bosons may be expected to involve the
 number operator itself under the square root sign. This is made exceedingly
 likely by analogy with the bose case, where the square root of the number
 operator in the zero momentum state appears. In Appendix B the manner 
 in which this exact
 correspondence may be deduced is hinted at. At this stage, it is 
 pertinent to merely write down a formula for
 the sea-boson annhilation operator
 in the RPA-limit.  The sea-boson
 is defined analogous to the condensate displacement boson, except
 the fermi case is more complicated due to the presence of the fermi surface.
 The sea-boson may be defined as follows (the rest of the details including
 a "proof" of this fact and how it fits into the fermi-bilinear-sea-boson
  correspondence is relegated to Appendix B),
\begin{equation}
a_{ {\bf{k}} }({\bf{q}}) = \frac{1}{\sqrt{ n_{ {\bf{k}} - {\bf{q}}/2 } }}
c^{\dagger}_{ {\bf{k}} - {\bf{q}}/2 }
(\frac{ n^{\beta}({\bf{k}} - {\bf{q}}/2) }{\langle N \rangle })^{\frac{1}{2}}
e^{i\mbox{ }\theta({\bf{k}}, {\bf{q}})}c_{ {\bf{k}} + {\bf{q}}/2 }
\label{SEABOSONFORM}
\end{equation}
 here $ \theta({\bf{k}}, {\bf{q}}) $ is a c-number phase that serves to 
 randomly cancel out troublesome terms :
 this is also related to the "random phase"
 of the Random Phase Approximation of Bohm and Pines. Thus the above formula
 for the sea-boson is in the "Random Phase" approximation.

 This correspondence recovers the salient features of the finite and zero 
 temperature aspects of the free theory provided we make the following
 assumption, the sea-bosons do participate in the thermodynamic averaging
 but come with an infinite negative chemical potential. This means that
 as far as the free theory is concerned, the average value of the sea-boson
 occupation is zero in the non-interacting case.
 The kinetic energy operator in the sea-boson language is given by,
\begin{equation}
K = \sum_{ {\bf{k}}, {\bf{q}} }\frac{ {\bf{k.q}} }{m}
\mbox{  }a^{\dagger}_{ {\bf{k}} }({\bf{q}})a_{ {\bf{k}} }({\bf{q}})
+ N\mbox{ }\epsilon_{0}
\end{equation}
here $ \epsilon_{0} $ is the kinetic energy per particle. Therefore,
\begin{equation}
\langle a^{\dagger}_{ {\bf{k}} }({\bf{q}}) a_{ {\bf{k}} }({\bf{q}})
\rangle = \frac{1}{exp(\beta(\frac{ {\bf{k.q}}}{m} - \mu_{B})) - 1}
 = 0
\end{equation}
 here $  - \mu_{B} = \infty $.
 However, when there are interactions in the system, the answer is likely
 to be different. In particular, it is likely to be a non-analytic function
 of the interaction in such a way that it vanishes as the coupling goes to
 zero(this is demonstrated explicitly in
 Appendix D). Roughly speaking we may write,
\begin{equation}
\langle a^{\dagger}_{ {\bf{k}} }({\bf{q}}) a_{ {\bf{k}} }({\bf{q}})
\rangle  \approx (\frac{1}{V})exp(-1/v)
\end{equation}
 where  $ v $ is the coulomb repulsion parameter and $ V $ is the volume.
 All these do come out
 naturally from the correspondece written down above as we shall soon see.
 We have thus written down a useful correspondence between fermi and
 bose operators that recovers the salient features of the free theory
 at zero and finite temperature and it is clear that this correspondence
 is all that is needed to write down model hamiltonians with any sort
 of interactions, like coulombic, with phonons e.t.c and extract exact
 non-perturbative (more precisely, non-analytic in the couplings) solutions.
 These solutions possess features that are impossible to capture via 
 diagrammatic means let alone mean-field theory. Thus a strong case is to
 be made for this method as a new paradigm for condensed matter physics.

\section{Field Operator in Terms of Density and its Canonical Conjugate}

 In this section, we introduce the canonical conjugate of the fermi/bose      
 density distribution. The reason for doing this is that we would like
 to express the field operator itself in terms of the density and its 
 canonical conjugate and consequently in terms of current and densities. 
 None of these ideas are really new. For example, Sunakawa et.al.
 \cite{Sun} use the velocity operator as a canonical conjugate of
 the density in their investigation of the properties of He-II.
 The velocity operator is somewhat related to the current operator but
 is not exactly equal to it. The reason is that the current operator 
 behaves like the conjugate of the density as far as commutation rules 
 with the latter is concerned, but does not commute with members 
 of its own kind(it is difficult to say this in words but will
 soon become clear). Let us postulate the existence of the object
 $ \Pi({\bf{x}}\sigma) $ as the canonical conjugate of the density,
\begin{equation}    
[\Pi({\bf{x}}\sigma), \rho({\bf{y}}\sigma^{'})] =
 i\mbox{ }\delta({\bf{x}} - {\bf{y}})\delta_{\sigma, \sigma^{'}}
\end{equation}
\begin{equation}
[\Pi({\bf{x}}\sigma), \Pi({\bf{y}}\sigma^{'})] = 0
\end{equation}
 It is clear that redefinitions of this object by amounts that involve
 translations by (more or less arbitrary) functionals
 of the density are not going to spoil the nature of the commutation
 rules above. However, we shall take the point of view that $ \Pi $ is
 defined to be that(almost unique) object that satisfies the relation below
 (making mathematically rigorous sense out of all this requires the
 use of functional analysis and will be attempted in Appendix E).
\begin{equation}
\rho({\bf{x}}\sigma) = -i \mbox{  }\frac{\delta}{\delta \Pi({\bf{x}}\sigma) }
\end{equation}
 Observe that
 $ \rho({\bf{x}}\sigma) = \psi^{\dagger}({\bf{x}}\sigma)\psi({\bf{x}}\sigma) $
 (technical problems involving the multiplication of operator-valued 
 distributions at the same point are alleviated by assuming
 that we have the whole system
 in a box, with periodic boundary conditions on the fields, making any 
 infinities only as large as the volume of the box itself, please refer
 to Appendix E for more details). Observe that (valid for
 both bose as well as fermi systems),
\begin{equation}
[\rho({\bf{x}}\sigma), \psi({\bf{x}}^{'}\sigma^{'})] = 
-\delta^{d}({\bf{x}}-{\bf{x}}^{'})\delta_{ \sigma, \sigma^{'} }
\psi({\bf{x}}\sigma)
\end{equation}
rewriting this as a differential equation,
\begin{equation}
[-i\frac{\delta}{\delta \Pi({\bf{x}}\sigma)}, \psi({\bf{x}}^{'}\sigma^{'})] =
-\delta^{d}({\bf{x}}-{\bf{x}}^{'})\delta_{ \sigma, \sigma^{'} }
\psi({\bf{x}}\sigma)
\end{equation}
This may be solved (exponentiation of commutation rules is the more
 technical term) as,
\begin{equation}
\psi({\bf{x}}\sigma) = exp(-i\Pi({\bf{x}}\sigma))F([\rho];{\bf{x}}\sigma)
\end{equation}
Observe now that $ \rho = \psi^{\dagger}\psi $. Therefore,
\begin{equation}
F^{\dagger}([\rho];{\bf{x}}\sigma)F([\rho];{\bf{x}}\sigma)
 = \rho({\bf{x}}\sigma)
\end{equation}
this may in turn be solved and the final
 Density Phase Variable Ansatz (DPVA for short) may be written as,
\begin{equation}
\psi({\bf{x}}\sigma) = e^{-i\mbox{ }\Pi({\bf{x}}\sigma)}
e^{i\mbox{ }\Phi([\rho];{\bf{x}}\sigma)}(\rho({\bf{x}}\sigma))^{\frac{1}{2}}
\label{DPVA}
\end{equation}
 It may be noted above that redefinitions of $ \Pi $ consistent with it being
 the canonical conjugate to $ \rho $ may be absorbed by a suitable redefinition
 of the phase functional $ \Phi $. Therefore, Eq.(~\ref{DPVA}) is in fact
 quite general.
 The crucial point of this whole exercise is that the phase functional 
 $ \Phi $ determines the statistics of the field $ \psi({\bf{x}}\sigma) $.
 It may be shown (the proof is rather tedious and since this issue is
 not central to the practical computations, we defer the proof for 
 a future communication) that imposition of bose/fermi
 commutation rules on $ \psi $ involves imposing the following
 restriction on the form of $ \Phi $.
\[
\Phi([\{\rho({\bf{y_{1}}}\sigma_{1})
 - \delta({\bf{y_{1}}}-{\bf{x}}^{'})\delta_{\sigma_{1},\sigma^{'}} \} ]
;{\bf{x}}\sigma)
\]
\[
+ \Phi([\rho];{\bf{x^{'}}}\sigma^{'}) - \Phi([\rho];{\bf{x}}\sigma)
\]
\begin{equation}
-\Phi([\{\rho({\bf{y_{1}}}\sigma_{1})
 - \delta({\bf{y_{1}}}-{\bf{x}})\delta_{\sigma_{1},\sigma} \} ]
;{\bf{x^{'}}}\sigma^{'})
 = m\pi
\label{recur}
\end{equation}
 where m is an odd integer for fermions even for bosons.
 This recursion is to be satisfied for all
 $ ({\bf{x}}\sigma) \neq  ({\bf{x^{'}}}\sigma^{'}) $. 
 It will be shown later that the restriction is far more severe, 
 brought about by the need to recover the free case properly. It may puzzle
 the reader that the above statement implies that a random choice of
 the phase functional that ensures that the recursion is satisfied 
 does not suffice. This is mysterious, but is clarified by a conjecture
 in Appendix E. This is done by relating the canonical conjugate to the
 current operator and rewriting the DPVA  in terms of current and
 densities. Again this type of idea has been addressed in the paper by 
 Goldin et. al.\cite{Sharp}. However, many in this field(pardon the pun!)
 continue to be
 under the mistaken impression that the formula for the annhilation operator
 (say the fermi operator) in terms of the corresponding currents and densities
 depends on whether the fields in question are free or whether 
 there are interactions in the system. This is shown to be false in the bose
 case, by demonstrating that there is a unique $ \Phi $ namely $ \Phi = 0 $
 that reproduces the free theory properly. Interactions just change
 the form of the hamiltonian but do not affect the form of the field operator
 in terms of currents and densities. The same is true but not easily seen
 in the fermi case; indeed throught this article we find that the bose case
 is much simpler and we shall take refuge under this rigorously justifiable 
 edifice when confronted by fermi systems. Further, the formulas for the
 field operators suggested by Goldin, Menikoff and Sharp in their 
 famous paper\cite{Sharp}
 are according to our results only partially correct, since they have
 not actually introduced the phase functional $ \Phi $ and computed it
 (this will again become clear soon).

 Let us now write down a formula for the current operator in terms
 of the canonical conjugate and density,
\begin{equation}
{\bf{J}} = (\frac{1}{2i})(\psi^{\dagger}\nabla\psi - (\nabla\psi)^{\dagger}\psi)
\end{equation}
using the DPVA Eq.(~\ref{DPVA}),
\begin{equation}
{\bf{J}}({\bf{x}}\sigma) = \rho(\nabla\Phi) 
- \rho(\nabla\Pi + [-i\Phi,\nabla\Pi])
\end{equation}
From this it possible to deduce a formula for the conjugate in terms 
 of currents and densities,
\begin{equation}
\Pi({\bf{x}}\sigma) = X_{ 0\sigma } + 
\int^{ {\bf{x}} }d{\bf{l}}\mbox{  }
(-1/\rho({\bf{y}}\sigma)){\bf{J}}({\bf{y}}\sigma)
 + \Phi([\rho];{\bf{x}}\sigma) 
 - \int^{ {\bf{x}} }d{\bf{l}}\mbox{  }[-i\Phi,\nabla\Pi]({\bf{y}}\sigma)
\end{equation}
 The line integral is along an arbitrary path from a remote point where
 all quantities may be set equal to zero. The field operator may now be 
 rewritten exclusively in terms of currents and densities like so,
\begin{equation}
\psi({\bf{x}}\sigma) = 
e^{-i\mbox{  }X_{0\sigma}-i\mbox{ }\int^{ {\bf{x}} }d{\bf{l}}\mbox{  }
(-1/\rho({\bf{y}}\sigma)){\bf{J}}({\bf{y}}\sigma)
 -i\mbox{  } \Phi([\rho];{\bf{x}}\sigma)
 + i\mbox{  } \int^{ {\bf{x}} }d{\bf{l}}\mbox{  }[-i\Phi,\nabla\Pi]({\bf{y}}\sigma) }
e^{i\mbox{  }\Phi([\rho];{\bf{x}}\sigma) }(\rho({\bf{x}}\sigma))^{\frac{1}{2}}
\label{FIELDOP}
\end{equation}
 here $ X_{0\sigma} $ is canonically conjugate to the total number
 of fermions/bosons($ [X_{0\sigma}, N_{\sigma^{'}}] = i\mbox{  }\delta_{\sigma,\sigma^{'}} $)
 $ N_{\sigma} = \sum_{ {\bf{k}} }c^{\dagger}_{ {\bf{k}}\sigma }c_{ {\bf{k}}\sigma } $
 The need for this is clear. The gradient of $ \Pi $ does not involve 
 the object $ X_{0\sigma} $ , when in fact it should. To put it differently,
 the field operator when commuted with $ N_{\sigma} $ should produce
 itself, whereas if we omit the object $ X_{0\sigma} $ then we find that
 the field operator commutes with the total number, which should not happen.
 These nuances are not very important for the practical computations
 as we shall see.
 It will be shown later that for bosons $ \Phi = 0 $  is the only possible
 choice and for fermions $ \Phi $ has to be fixed by making contact with the
 free theory. Uniqueness is assumed for the fermi case by making analogy with
 the bose case for which uniqueness may be proved.
 
\section{Making Contact with the Free Theory}

 In this section, we write down the kinetic energy operator in terms of
 the sea-displacements and determine the undetermined phase functional 
 $ \Phi $ in the fermi case. The reason why the phase functional 
 $ \Phi = 0 $ in the bose case will also be addressed here.
 Let us take the bose case first. It is clear at the outset that 
 the choice $ \Phi = 0 $ satisfies the recursion Eq.(~\ref{recur}) 
 for bosons when
 one assumes that $ m = 0 $, an even integer. That this is the only
 possible choice is not at all clear. In order to verify this,
 let us write down the kinetic energy operator in terms of the density
 and its conjugate and show that an expansion in terms of the density
 fluctuations recovers the correct form of the dynamical density
 correlation function of the free theory(just the bose case).
\begin{equation}
K = \int\frac{ d{\bf{x}} }{2m}\mbox{  }[ \rho(\nabla\Pi)^{2} 
+ \frac{(\nabla\rho)^{2}}{4\rho} ] + c-number
\end{equation}
 It may now be verified that an expansion in terms of density fluctuations 
 leads to a hamiltonian that describes free harmonic oscillators, that
 may be easily diagonalised, it may also be shown that this diagonalised
 form reproduces the correct dynamical density correlation functions.
 The expanded form of the operator in fourier space is reproduced below for
 convenience,
\begin{equation}
K = \sum_{ {\bf{q}} \neq 0 }N\mbox{  }\epsilon_{ {\bf{q}} }X_{ {\bf{q}} }
X_{ -{\bf{q}} }
 + \sum_{ {\bf{q}} \neq 0 }
\frac{ \epsilon_{ {\bf{q}} } }{4N}\rho_{ {\bf{q}} }\rho_{ -{\bf{q}} }
\end{equation}
 A different choice of $ \Phi $ does not reproduce the free theory correctly.
 This is attested to by a simple calculation made in 1D. Let us assume
 a form,
\begin{equation}
\Phi([\rho];x) = 2\pi\int_{-\infty}^{+\infty}\mbox{ }dy\mbox{ }
\theta(x-y)(\rho(y)-\rho_{0})
\end{equation}
where $ \theta(x) $ is the Heaviside step function.
The above form clearly satisfies the recursion but does not reproduce the 
 free theory as may be easily verified by the reader.  

 The fermi case is somewhat more difficult. The difficulty is due to the
 fact that we must have a choice of $ \Phi \neq 0 $
 that satisifes the recursion
 at the same time reproducing the free case. We shall take the point of view
 that the simplest choice for $ \Phi $ namely linear in $ \rho $ should 
 suffice. In any event, for the scheme to have practical significance, it is
 important for $ \Phi $ to be a simple functional of the density. 
 We fix the coefficient in this ansatz by making contact with the free theory.
 Let us focus on the case of spinless fermions.  In what follows, we restrict 
 ourselves to zero temperature and a weakly non-ideal system, in this case, 
 we are allowed to
 replace the $ {\bar{n}}_{ {\bf{k}} } $ in the definition of 
 $ \Lambda_{ {\bf{k}} }({\bf{q}}) $ by its noninteracting value at zero
 temperature. More interesting situations arise when the quantity 
 $ {\bar{n}}_{ {\bf{k}} } $ is evaluated self-consistently, but we shall 
 relegate these issues to future publications \cite{Setlur}.
 From equation Eq.(~\ref{FIELDOP}) it is clear that redefinitions
 of the phase functional by amounts that do not depend on the density
 for example,
 $ \Phi([\rho];{\bf{x}}) \rightarrow \Phi([\rho];{\bf{x}}) + f({\bf{x}}) $,
 do not affect the formula for the field operator in Eq.(~\ref{FIELDOP}).
 Therefore, let us try the following ansatz for $ \Phi $.
\begin{equation}
\Phi([\rho];{\bf{x}}) = \sum_{ {\bf{q}} \neq 0 }U_{ {\bf{q}} }({\bf{x}})
\rho_{ {\bf{q}} }
\end{equation}
Let us now write down the kinetic energy operator for fermions
 using the results of the first section. The kinetic energy operator
 was written down as,
\begin{equation}
K = \sum_{ {\bf{k}}, {\bf{q}} }\frac{ {\bf{k.q}} }{m}
a^{\dagger}_{ {\bf{k}} }({\bf{q}})
a_{ {\bf{k}} }({\bf{q}}) + N \mbox{ }\epsilon_{0}
\end{equation}
 It has been demonstrated in Appendix C that if one uses the form of the
 density fluctuation operator obtained by dropping quadratic terms in the
 sea displacements(the existence of such quadratic terms are hinted
 in Appendix B), this reproduces the RPA dielectric function. Since we
 know from prior experience that RPA is exact in the ultra-high density limit,
 we can use these two pieces of information  to deduce a formula for
 $ U_{ {\bf{q}} }({\bf{x}}) $ in terms of the properties of the free theory.
First let us write down the RPA-form of the density fluctuation operator,
\begin{equation}
{\tilde{\rho}}_{ {\bf{q}} } = \sum_{ {\bf{k}} }
[\Lambda_{ {\bf{k}} }({\bf{q}}) a_{ {\bf{k}} }(-{\bf{q}})
 + \Lambda_{ {\bf{k}} }(-{\bf{q}}) a^{\dagger}_{ {\bf{k}} }({\bf{q}})]
\end{equation}
where,
\begin{equation}
\Lambda_{ {\bf{k}} }({\bf{q}}) = 
\sqrt{ {\bar{n}}_{ {\bf{k}} + {\bf{q}}/2 }
(1 - {\bar{n}}_{ {\bf{k}} - {\bf{q}}/2 }) }
\end{equation}
and the corresponding conjugate variable may be written down
 (that is, $ \Pi $ in fourier space),
\begin{equation}
{\tilde{X}}_{ {\bf{q}} } = 
(-\frac{1}{2\mbox{ }i\mbox{ }N\mbox{ }\epsilon_{ {\bf{q}} }})
\sum_{ {\bf{k}} }
[\Lambda_{ {\bf{k}} }(-{\bf{q}}) \omega_{ {\bf{k}} }({\bf{q}})
a_{ {\bf{k}} }({\bf{q}})
 - \Lambda_{ {\bf{k}} }({\bf{q}})\omega_{ {\bf{k}} }(-{\bf{q}})
 a^{\dagger}_{ {\bf{k}} }(-{\bf{q}})]
\end{equation}
the dispersion is given by
 $ \omega_{ {\bf{k}} }({\bf{q}}) = \frac{ {\bf{k.q}} }{m} $.
From this the fermi field operator may be written down as,
\begin{equation}
\psi({\bf{x}}) = e^{-i\mbox{ }U_{1}({\bf{x}}) }e^{i\mbox{ }U_{2}({\bf{x}})}
\sqrt{\rho_{0}}
\end{equation}
where,
\begin{equation}
U_{1}({\bf{x}}) = \sum_{ {\bf{q}} \neq 0 }e^{i\mbox{ }{\bf{q.x}} }
{\tilde{X}}_{ {\bf{q}} }
\end{equation}
\begin{equation}
U_{2}({\bf{x}}) = \sum_{ {\bf{q}} \neq 0 }U_{ {\bf{q}} }({\bf{x}})
{\tilde{\rho}}_{ {\bf{q}} }
\end{equation}
 Using these facts, let us compute the equal-time version of the propagator
 below in the bose language and in the usual fermi language and equate the two
 expressions. In the sea-displacement language it comes out as,
\begin{equation}
\langle \psi^{\dagger}({\bf{x}},t)\psi({\bf{x}}^{'},t)\rangle = 
\rho_{0}\mbox{  }e^{-\sum_{ {\bf{k}}, {\bf{q}} \neq 0}
g^{*}_{ {\bf{k}}, {\bf{q}} }({\bf{x}})g_{ {\bf{k}}, {\bf{q}} }({\bf{x}}) }
e^{\sum_{ {\bf{k}}, {\bf{q}} \neq 0}g^{*}_{ {\bf{k}}, {\bf{q}} }({\bf{x}})
g_{ {\bf{k}}, {\bf{q}} }({\bf{x}}^{'}) }
\end{equation}
where,
\begin{equation}
g_{ {\bf{k}}, {\bf{q}} }({\bf{x}}) = -e^{-i\mbox{ }{\bf{q.x}} }
(\frac{1}{2\mbox{ }N\mbox{ }\epsilon_{ {\bf{q}} } })
\Lambda_{ {\bf{k}} }(-{\bf{q}})\omega_{ {\bf{k}} }({\bf{q}})
 + i\mbox{ }U_{ {\bf{q}} }({\bf{x}})\Lambda_{ {\bf{k}} }(-{\bf{q}})
= -f^{*}_{ {\bf{k}}, {\bf{q}} }({\bf{x}})
\end{equation}
In the original fermi language it is,
\begin{equation}
\langle \psi^{\dagger}({\bf{x}},t)\psi({\bf{x}}^{'},t)\rangle =
\frac{1}{V}\sum_{ {\bf{q}} }e^{i\mbox{ }{\bf{q}}.({\bf{x}}^{'} - {\bf{x}})}
\theta(k_{f} - |{\bf{q}}|)
\end{equation}
Set $ U_{ {\bf{q}} }({\bf{x}}) = e^{-i\mbox{ }{\bf{q.x}}}U_{0}({\bf{q}}) $
 and $ U_{0}({\bf{q}}) $ is real.
In order to derive a formula for $ U_{0}({\bf{q}}) $, let us equate the
 logarithm of the two expressions,
\[
log(\langle \psi^{\dagger}({\bf{x}},t)\psi({\bf{x}}^{'},t)\rangle)
 = log(\rho_{0}) + \sum_{ {\bf{k}}, {\bf{q}} \neq 0 }
[(\frac{1}{2\mbox{ }N\mbox{ }\epsilon_{ {\bf{q}} } })^{2}
(\frac{ {\bf{k.q}} }{m})^{2} + (U_{0}({\bf{q}}))^{2}]
(\Lambda_{ {\bf{k}} }(-{\bf{q}}))^{2}(e^{i{\bf{q}}.({\bf{x}}-{\bf{x}}^{'})} - 1)
\]
\begin{equation}
 = log(\rho_{0}) + log(1 + \frac{1}{N}\sum_{ {\bf{q}} \neq 0 }
(e^{i{\bf{q}}.({\bf{x}}-{\bf{x}}^{'})} - 1)\theta(k_{f} - |{\bf{q}}|))
 \approx log(\rho_{0}) + \frac{1}{N}\sum_{ {\bf{q}} \neq 0 }
(e^{i{\bf{q}}.({\bf{x}}-{\bf{x}}^{'})} - 1)\theta(k_{f} - |{\bf{q}}|)
\end{equation}
This leads to the following formula for the coefficient,
\begin{equation}
U_{0}({\bf{q}}) = \frac{1}{N}
(\frac{ \theta(k_{f} - |{\bf{q}}|) - w_{1}({\bf{q}}) }{ w_{2}({\bf{q}}) })
^{\frac{1}{2}}
\end{equation}
\begin{equation}
w_{1}({\bf{q}}) = (\frac{1}{4\mbox{ }N\mbox{ }\epsilon^{2}_{ {\bf{q}} }})
\sum_{ {\bf{k}} }(\frac{ {\bf{k.q}} }{m})^{2} 
(\Lambda_{ {\bf{k}} }(-{\bf{q}}))^{2}
\end{equation}
\begin{equation}
w_{2}({\bf{q}}) = (\frac{1}{N})\sum_{ {\bf{k}} }
(\Lambda_{ {\bf{k}} }(-{\bf{q}}))^{2}
\end{equation}
 In fact, in principle we could 
 go all the way back to the expression in Eq.(~\ref{FIELDOP}) and say
 that we now have a unique correspondence between the fermi field operator
 and the corresponding currents and densities. In the next section, 
 we write down and diagonalise the hamiltonian of interacting systems.
 It is shown that when only the lowest order sea-displacement terms/condensate
 displacement terms are included, it amounts to doing RPA/Bogoliubov theory. 
 This hamiltonian is diagonalised in the fermi and bose cases and the
 single-particle spectral functions are computed. The bose case comes out 
 nicely since, it is just the bogoliubov theory, but in the fermi case, we
 have to take extra care in properly diagonalising the hamiltonian in order
 not to lose the particle hole mode, the collective mode being more obvious.
 
\section{Spectral Function of Interacting Systems}
 
Let us make the following observation for future reference,
\begin{equation}
{\bf{RPA/Bogoliubov \rightarrow Leave \mbox{  }out\mbox{  }the
\mbox{  } Quadratic \mbox{  }Part\mbox{  } in\mbox{  } 
Eq.(~\ref{FERMI})\mbox{  } and \mbox{  }Eq.(~\ref{BOSE}) }}
\label{PRES}
\end{equation}
 It is pertinent at this stage to remark on the physical meaning of the
 above relation. In case of bosons, it is simple to visualise.
 Bogoliubov's theory is exact provided there are large number bosons
 in the zero momentum state so that we may legitimately replace the 
 number operator by its c-number expectation value. Also it is important that
 the system be weakly interacting so that the fluctuations of the number
 operator in the zero momentum state are small compared with its macroscopic 
 expectation value. In the fermi case an analogous statement would be that 
 the momentum distribution be sufficiently different from zero or unity 
 for all values of the momenta. Also the fluctuations of the momentum 
 distribution must be small. Thus for the fermi system, our approach gives 
 good answers even for strong interactions that drive the momentum
 distribution away from zero or unity for
 all momenta so long as the fluctuations around these nonideal averages
 are small. In any event, the philosophy is that we have an exactly solvable
 class of models that describe correlation effects in many different
 contexts and this alone merits attention and serious study. In the end 
 experiments may have to be used to 'calibrate' these models so that they
 become a true description of the low-energy real world. 

\begin{center} {\bf{Part A}} \end{center}
 Let us focus on the bose case first. The bogoliubov hamiltonian
 may be written down by following the prescription of Eq.(~\ref{PRES}).
\begin{equation}
H_{bog} = \sum_{ {\bf{k}} }\epsilon_{ {\bf{k}} }
d^{\dagger}_{ (1/2){\bf{k}} }({\bf{k}})d_{ (1/2){\bf{k}} }({\bf{k}})
+ \sum_{ {\bf{q}} \neq 0 }\frac{ v_{ {\bf{q}} } }{2}
[{\sqrt{N_{0}}}d_{ -{\bf{q}}/2 }(-{\bf{q}})
+ d^{\dagger}_{ {\bf{q}}/2 }({\bf{q}}){\sqrt{N_{0}}}]
[{\sqrt{N_{0}}}d_{ {\bf{q}}/2 }({\bf{q}})
+ d^{\dagger}_{ -{\bf{q}}/2 }(-{\bf{q}}){\sqrt{N_{0}}}]
\end{equation}
 In the above equation $ N_{0} $ is an operator,
 therefore this is the non-local
 bogoliubov hamiltonian. But we shall assume that it is legitimate to replace
 it with its c-number expectation value. It would be interesting to see
 what corrections to
 bogoliubov theory come about by incorporating this square-root of the
 operator. These correction terms tell us that fluctuations of the number of
 particles in the condensate are important and lead to correlations beyond the
 bogoliubov theory. This is in addition to correlations coming from quadratic
 terms that the prescription  Eq.(~\ref{PRES}) neglects. When these 
 approximations are implemented, 
 and a further approximation $ N_{0} \approx N  $ is made, it becomes
 exactly the bogoliubov theory introduced by Bogoliubov 
 and Bogliubov and Zubarev\cite{Bog}. It may be diagonalised quite easily,
\begin{equation}
H_{bog} = \sum_{ {\bf{q}} }\omega_{ {\bf{q}} }
f^{\dagger}_{ {\bf{q}} }f_{ {\bf{q}} }
\end{equation}
and,
\begin{equation}
f_{ {\bf{q}} } =
 (\frac{ \omega_{ {\bf{q}} } + \epsilon_{ {\bf{q}} }
 + \rho_{0}v_{ {\bf{q}} } }
{ 2 \mbox{ }\omega_{ {\bf{q}} } })^{\frac{1}{2}}
d_{ {\bf{q}}/2 }({\bf{q}})
 + (\frac{ -\omega_{ {\bf{q}} } +
 \epsilon_{ {\bf{q}} } + \rho_{0}v_{ {\bf{q}} } }
{ 2 \mbox{ }\omega_{ {\bf{q}} } })^{\frac{1}{2}}
d^{\dagger}_{ -{\bf{q}}/2 }(-{\bf{q}})
\end{equation}
\begin{equation}
d_{ {\bf{q}}/2 }({\bf{q}})
 = (\frac{ \omega_{ {\bf{q}} } + \epsilon_{ {\bf{q}} }
 + \rho_{0}v_{ {\bf{q}} } }
{ 2 \mbox{ }\omega_{ {\bf{q}} } })^{\frac{1}{2}}f_{ {\bf{q}} }
 - (\frac{ -\omega_{ {\bf{q}} } + \epsilon_{ {\bf{q}} }
 + \rho_{0}v_{ {\bf{q}} } }
{ 2 \mbox{ }\omega_{ {\bf{q}} } })^{\frac{1}{2}}f^{\dagger}_{ -{\bf{q}} }
\end{equation}
 The dispersion is given by,
\begin{equation}
\omega_{ {\bf{q}} } = \sqrt{ \epsilon^{2}_{ {\bf{q}} }
+ 2 \rho_{0}v_{ {\bf{q}} }\epsilon_{ {\bf{q}} } }
\end{equation}
 here, $ \rho_{0} $ is the density of bosons in the condensate(not the overall
 density). From this one may deduce the filling fraction and dynamical
 structure factor,
\newline
{\center{ {\bf{FILLING FRACTION}} }}
\begin{equation}
f_{0} = N_{0}/N = 1 -
(1/N)\sum_{ {\bf{q}} }\langle d^{\dagger}_{ (1/2){\bf{q}} }({\bf{q}})
d_{ (1/2){\bf{q}} }({\bf{q}}) \rangle
\end{equation}
in other words,
\begin{equation}
f_{0} = N_{0}/N = 1 - (1/2\pi^{2}\rho)\int_{0}^{\infty}
 \mbox{ }dq\mbox{ }
q^{2} ( \frac{ -\omega_{ {\bf{q}} } + \epsilon_{ {\bf{q}} }
+ \rho_{0}v_{ {\bf{q}} } }{ 2\mbox{ }\omega_{ {\bf{q}} } } )
\end{equation}
here $ \rho $ is the total density of bosons including those that are
 not in the condensate.
\newline
{\bf{DYNAMICAL STRUCTURE FACTOR}}
\begin{equation}
S^{>}({\bf{q}}t) = \langle \rho_{ {\bf{q}} }(t) \rho_{ -{\bf{q}} }(0)\rangle
 = N_{0}\langle [ d_{ -(1/2){\bf{q}} }(-{\bf{q}})(t)
 + d^{\dagger}_{ (1/2){\bf{q}} }({\bf{q}})(t) ]
[ d_{ (1/2){\bf{q}} }({\bf{q}})(0)
 + d^{\dagger}_{ -(1/2){\bf{q}} }(-{\bf{q}})(0) ] \rangle
\end{equation}
in other words,
\begin{equation}
S^{>}({\bf{q}},t) = N_{0}
(\frac{ \epsilon_{ {\bf{q}} } }{\omega_{ {\bf{q}} } })
 \mbox{ }exp(-i \mbox{ }\omega_{ {\bf{q}} }t)
\end{equation}
 This method is truly powerful when applied to compute single-particle
 properties. The single-particle green function is difficult to obtain
 using conventional digrammatic methods or otherwise
 (See Kadanoff and Baym Ref.(~\onlinecite{Baym}) ).
 For this one must first write
 down the  field operator in terms of the condensate-displacements.
\begin{equation}
\Pi({\bf{x}}) \approx  (\frac{i}{2\sqrt{N_{0}}})
\sum_{ {\bf{q}} }exp(i{\bf{q.x}})[d_{ {\bf{q}}/2 }({\bf{q}})
 - d^{\dagger}_{ -{\bf{q}}/2 }(-{\bf{q}})]
\end{equation}
and the expression for the field operator is,
\begin{equation}
\psi({\bf{x}}) \approx e^{-i\mbox{ }\Pi({\bf{x}}) } \mbox{ }
\sqrt{ \rho }
\end{equation}
 The propagator (all propagators in this article are evaluated at
 zero temperature, this means we may set the
 chemical potential equal to zero in the bose case)
 may now be computed and shown to be equal to the
 free propagator at ultra-high density. The interacting case is more
 interesting.
 The time-evolved version is,
\begin{equation}
\psi({\bf{x}},t) \approx e^{-i\mbox{ }\Pi({\bf{x}},t) } \mbox{ }
\sqrt{ \rho }
\end{equation}
and
\begin{equation}
\Pi({\bf{x}},t) = (\frac{i}{2\sqrt{N_{0}}})
\sum_{ {\bf{q}} }exp(i{\bf{q.x}})
(A_{ {\bf{q}} }+B_{ {\bf{q}} })
[f_{ {\bf{q}} }
e^{-i\omega_{ {\bf{q}} }t }
 - f^{\dagger}_{ -{\bf{q}} }e^{i\omega_{ {\bf{q}} }t }]
\end{equation}
\begin{equation}
\langle \psi^{\dagger}({\bf{0}},0)\psi({\bf{x}},t)\rangle
 = \rho\mbox{   }\langle \mbox{ }e^{i\mbox{  }\Pi({\bf{0}},0)}
e^{-i\mbox{  }\Pi({\bf{x}},t)} \mbox{ }\rangle
\end{equation}
 In order to ensure that the free case is properly recovered, we use this
 somewhat illegal trick, but a trick that should be very palatable to
 most physicists, namely multiply and divide by the free propagator and
 in the division use the free propagator predicted by the bosonized theory and
 in the numerator use the free propagator obtained from elementary
 considerations.
\begin{equation}
\langle \psi^{\dagger}({\bf{0}},0)\psi({\bf{x}},t)\rangle
 = exp\mbox{ }[ \mbox{  }(\frac{1}{4N_{0}})
\sum_{ {\bf{q}} }f_{ {\bf{q}} }({\bf{x}},t) \mbox{  }]
\langle \psi^{\dagger}({\bf{0}},0)\psi({\bf{x}},t)\rangle_{free}
\end{equation}
where,
\begin{equation}
A_{ {\bf{q}} } = (\frac{ \omega_{ {\bf{q}} } + \epsilon_{ {\bf{q}} }
+ \rho_{0}v_{ {\bf{q}} } }{2 \mbox{ }\omega_{ {\bf{q}} } })^{\frac{1}{2}}
\end{equation}
\begin{equation}
B_{ {\bf{q}} } = (\frac{ -\omega_{ {\bf{q}} } + \epsilon_{ {\bf{q}} }
+ \rho_{0}v_{ {\bf{q}} } }{2 \mbox{ }\omega_{ {\bf{q}} } })^{\frac{1}{2}}
\end{equation}
Similarly,
\[
\langle \psi({\bf{x}},t)\psi^{\dagger}({\bf{0}},0) \rangle
 = \rho\langle e^{-i\mbox{ }\Pi({\bf{x}},t)}
e^{i\mbox{ }\Pi({\bf{0}},0)} \rangle
\]
\begin{equation}
= exp\mbox{ }[ \mbox{  }(\frac{1}{4N_{0}})
\sum_{ {\bf{q}} }f_{ {\bf{q}} }(-{\bf{x}},-t) \mbox{  }]
\langle \psi({\bf{x}},t)\psi^{\dagger}({\bf{0}},0)\rangle_{free}
\end{equation}
here,
\begin{equation}
f_{ {\bf{q}} }({\bf{x}},t) =
(e^{-i\mbox{ }{\bf{q.x}}}e^{i\mbox{ }\omega_{ {\bf{q}} }t}
 -1)(A_{ {\bf{q}} } + B_{ {\bf{q}} })^{2}
 - (e^{-i\mbox{ }{\bf{q.x}}}e^{i\mbox{ }\epsilon_{ {\bf{q}} }t}
 -1)
\end{equation}
From Kadanoff and Baym \cite{Baym} the spectral function may be deduced
 as follows,
\newline
{\bf{THE SPECTRAL FUNCTION}}
\begin{equation}
A({\bf{p}},\omega) = \int \mbox{  }{ d{\bf{x}} }\mbox{  }
\int_{ -\infty }^{ +\infty }\mbox{ }dt\mbox{ }
e^{-i{\bf{p.x}} + i\mbox{ }\omega\mbox{ }t}
\{ exp[\frac{1}{4N_{0}}\sum_{ {\bf{q}} }f_{ {\bf{q}} }(-{\bf{x}},-t) ]
(\rho + u({\bf{x}},t))
 -  exp[\frac{1}{4N_{0}}\sum_{ {\bf{q}} }f_{ {\bf{q}} }({\bf{x}},t) ]
\rho \}
\end{equation}
and,
\begin{equation}
u({\bf{x}},t) = \frac{1}{V}\sum_{ {\bf{k}} }e^{i\mbox{ }{\bf{k.x}} }
e^{-i\mbox{ }\epsilon_{ {\bf{k}} }t }
\end{equation}
 The above answer is the exact answer for the spectral function provided 
 bogoliubov's theory is adequate. Now we move on to the fermi case which is
 far more interesting and important.
\begin{center} {\bf{Part B}} \end{center}
 In order to compute the full propagator for these systems, it is desirable
 to first ascertain, under what conditions these formulas are going to be
 valid. The answer is given by the assertion in Eq.(~\ref{PRES}). Thus
 these answers for the single-particle properties are valid in the same limit
 in which RPA/Bogoliubov's theory is exact. The assertion in the bose case
 in Eq.(~\ref{PRES}) has been verified.
 In order to verify the analogous assertion in the fermi case, we have
 to diagonalise the full hamiltonian given below (The fact that the RPA
 dielectric function comes out naturally from the prescription in 
 Eq.(~\ref{PRES}) will be demsonstrated in Appendix C).
 In the fermi case, we have to diagonalise the full hamiltonian given below,
\begin{equation}
H = \sum_{ {\bf{k}}, {\bf{q}} }\omega_{ {\bf{k}} }({\bf{q}})
a^{\dagger}_{ {\bf{k}} }({\bf{q}})a_{ {\bf{k}} }({\bf{q}})
 + \sum_{ {\bf{q}} \neq 0 }\frac{ v_{ {\bf{q}} } }{2V}
\sum_{ {\bf{k}}, {\bf{k}}^{'} }
[\Lambda_{ {\bf{k}} }({\bf{q}})a_{ {\bf{k}} }(-{\bf{q}})
 + \Lambda_{ {\bf{k}} }(-{\bf{q}})a^{\dagger}_{ {\bf{k}} }({\bf{q}})]
[\Lambda_{ {\bf{k}}^{'} }(-{\bf{q}})a_{ {\bf{k}}^{'} }({\bf{q}})
 + \Lambda_{ {\bf{k}}^{'} }({\bf{q}})a^{\dagger}_{ {\bf{k}}^{'} }(-{\bf{q}})]
\label{FULLHAM}
\end{equation}
 here
$ \omega_{ {\bf{k}} }({\bf{q}}) =
 \frac{ {\bf{k.q}} }{m}\Lambda_{ {\bf{k}} }(-{\bf{q}}) $.
 The zero temperature case is somewhat special, here we may assume
 that the sea-boson annhilates the non-interacting fermi sea, this means
 that we have to introduce a factor of $ \Lambda_{ {\bf{k}} }(-{\bf{q}}) $
 in the dispersion that makes the kinetic energy operator positive
 definite.
 In order to diagonalise this we proceed as follows. Assume that the
 diagonalised form looks like this,
\begin{equation}
H = \sum_{ i, {\bf{q}} }{\tilde{\omega}}_{ i }({\bf{q}})
b^{\dagger}_{ i }({\bf{q}})b_{ i }({\bf{q}})
\end{equation}
 here $ b_{ i }({\bf{q}}) $ and $ b^{\dagger}_{ i }({\bf{q}}) $
 are "dressed-sea-displacement" operators. The objects $ i $ take on 
 values from an index set. The size of this set is the big issue here. Is it
 finite or does it have the same size as the number of points in k-space,
 or is it equal to the number of points on the fermi surface ?  We shall find
 that answers to these questions are hard, and may be addressed only after
 coming to an agreement as to what sort of physics we hope to capture. 
 Indeed, in many cases in physics one is forced to bend the rules or reinterpret
 mathematical formulae in order to capture what one is looking for. We find
 that we have to resort to such methods here as well. In particular, we find
 the following general feature. $ {\tilde{\omega}}_{i}({\bf{q}}) $
 are the  roots
 of the RPA-dielectric function. Now the RPA-dielectric function is a complex
 quantity, as it is usually introduced in the textbooks. Therefore finding
 roots cannot mean finding the zeros of both the real and imaginary parts
 at the same time for this gives no root, both the real and imaginary
 part cannot be zero simultaneously. This leaves us with the following options,
 reinterpret the zeros of the RPA-dielectric function to be the maxima of the
 dynamical structure factor, in which case one gets both the particle-hole mode
 as well as the collective mode. The better option is to delay taking the 
 thermodynamic limit until after all the summation over momenta 
 have been performed. Then assume that the density is high enough and
 at the very end go to the thermodynamic limit, this ensures that both the
 particle-hole mode and the collective mode are properly recovered.  These 
 are admitedly difficult issues to grapple with, and the authors have
 attempted
 a different approach to deal with them. However, the traditional viewpoint
 on this matter is presented in the paper by Castro-Neto and Fradkin\cite{Neto}.
 The diagonalisation proceeds as follows,
\begin{equation}
b_{i}({\bf{q}}) = \sum_{ {\bf{k}} }
[b_{i}({\bf{q}}), a^{\dagger}_{ {\bf{k}} }({\bf{q}})]
a_{ {\bf{k}} }({\bf{q}})
 -  \sum_{ {\bf{k}} }
[b_{i}({\bf{q}}), a_{ {\bf{k}} }(-{\bf{q}})]
a^{\dagger}_{ {\bf{k}} }(-{\bf{q}})
\end{equation}
the corresponding inverted formula reads as,
\begin{equation}
a_{ {\bf{k}} }({\bf{q}}) = \sum_{ i }
[a_{  {\bf{k}} }({\bf{q}}), b^{\dagger}_{ i }({\bf{q}})]
b_{ i }({\bf{q}})
 -  \sum_{ i }
[a_{ {\bf{k}} }({\bf{q}}), b_{ i }(-{\bf{q}})]
b^{\dagger}_{ i }(-{\bf{q}})
\end{equation}
The quantities $ [b_{i}({\bf{q}}), a_{ {\bf{k}} }(-{\bf{q}})] $ and
 $ [a_{  {\bf{k}} }({\bf{q}}), b^{\dagger}_{ i }({\bf{q}})] $
are c-numbers and real. The $ i $ here could span a continuum
(particle-hole mode) or be finite( actually there is just one, 
 collective mode ). The diagonalisation continues unabated,
\begin{equation}
[b_{i}({\bf{q}}), a^{\dagger}_{ {\bf{k}} }({\bf{q}})]
 = (\frac{ \Lambda_{ {\bf{k}} }(-{\bf{q}}) }
{ {\tilde{\omega}}_{i}({\bf{q}}) - \frac{ {\bf{k.q}} }{m} })
g_{i}({\bf{q}}) = [a_{ {\bf{k}} }({\bf{q}}), b^{\dagger}_{i}({\bf{q}})]
\end{equation}
\begin{equation}
[b_{i}({\bf{q}}), a_{ {\bf{k}} }(-{\bf{q}})]
 = -(\frac{ \Lambda_{ {\bf{k}} }({\bf{q}}) }
{ {\tilde{\omega}}_{i}({\bf{q}}) - \frac{ {\bf{k.q}} }{m} })
g_{i}({\bf{q}})
\end{equation}
\begin{equation}
[a_{ {\bf{k}} }({\bf{q}}), b_{i}(-{\bf{q}})]
 = (\frac{ \Lambda_{ {\bf{k}} }(-{\bf{q}}) }
{ {\tilde{\omega}}_{i}(-{\bf{q}}) + \frac{ {\bf{k.q}} }{m} })
g_{i}(-{\bf{q}})
\end{equation}
\begin{equation}
g_{i}({\bf{q}}) =  [\sum_{ {\bf{k}} }
\frac{ n_{F}({\bf{k-q/2}}) - n_{F}({\bf{k+q/2}}) }
{ ( {\tilde{\omega}}_{i}({\bf{q}}) - \frac{ {\bf{k.q}} }{m} )^{2} }]
^{-\frac{1}{2}}
\end{equation}
The eigenvalues $ {\tilde{\omega}}_{i}({\bf{q}}) $ are zeros of the
 real part of the RPA dielectric function.
 The RPA dielectric function is written down below,
\begin{equation}
\epsilon_{RPA}({\bf{q}}, {\tilde{\omega}}) = 1 + \frac{ v_{ {\bf{q}} } }{V}
\sum_{ {\bf{k}} }
\frac{ n_{F}({\bf{k}} + {\bf{q}}/2) - n_{F}({\bf{k}} - {\bf{q}}/2) }
{ {\tilde{\omega}} - \frac{ {\bf{k.q}} }{m} }
\label{RPA}
\end{equation}  
 As it stands, the above sum is ill-defined. In particular, if one takes
 the thermodynamic limit at the outset, and treats the above sum as the
 principal part, then one gets the real part of the RPA dielectric function. 
 On the other hand, if one defers the taking of the thermodynamic limit
 until the very end, and instead takes the high density limit first,
 then one obtains the particle-hole mode as the argument below will attest.
 Let us rewrite the sum in the RPA dielectric function as,
\[
\epsilon_{RPA}({\bf{q}}, {\tilde{\omega}}) = 1 + \frac{ v_{ {\bf{q}} } }{V}
\sum_{ {\bf{k}} }
\frac{ \Lambda^{2}_{ {\bf{k}} }({\bf{q}}) -  \Lambda^{2}
_{ {\bf{k}} }(-{\bf{q}}) }
{ {\tilde{\omega}} - \frac{ {\bf{k.q}} }{m} }
\]
\[
 = 1 + \frac{ v_{ {\bf{q}} } }{V}
\sum_{ {\bf{k}} }\frac{ \Lambda^{2}_{ {\bf{k}} }({\bf{q}}) }
{ {\tilde{\omega}} + \omega_{ {\bf{k}} }(-{\bf{q}}) }
- \frac{ v_{ {\bf{q}} } }{V}
\sum_{ {\bf{k}} \neq {\bf{k}}_{i}}\frac{ \Lambda^{2}_{ {\bf{k}} }(-{\bf{q}}) }
{ {\tilde{\omega}} - \omega_{ {\bf{k}} }({\bf{q}}) }
\]
\begin{equation}
- \frac{ v_{ {\bf{q}} } }{V}
\frac{ \Lambda^{2}_{ {\bf{k}}_{i} }(-{\bf{q}}) }
{ {\tilde{\omega}} - \omega_{ {\bf{k}}_{i} }({\bf{q}}) }
\label{REDU}
\end{equation}                                           
 Let us now assume that the volume $ V $ is fixed and we now go to the
 high density limit($ k_{F} \rightarrow \infty $, or equivalently when ,
 $ |{\bf{q}}| << k_{f} $),
 then we find, due to the fact below,
\begin{equation}
\Lambda_{ {\bf{k}} }(-{\bf{q}}) =0 \mbox{   }; \mbox{   }unless 
\mbox{   }|{\bf{k}}| \approx k_{f} \mbox{   }and  \mbox{   }
{\bf{k}}.{\bf{q}} > 0
\end{equation}
 The total number of terms in the above two sums is a small fraction of the
 total volume and as $  k_{F} $ keeps increasing, the fraction gets
 smaller and smaller until it becomes small compared to unity and may
 be neglected, this means,
\begin{equation}
1 - (\frac{ v_{ {\bf{q}} } }{V})
\frac{ \Lambda^{2}_{ {\bf{k}}_{i} }(-{\bf{q}}) }
{ {\tilde{\omega}}_{i}({\bf{q}}) - \omega_{ {\bf{k}}_{i} }({\bf{q}}) }
 = 0 
\end{equation}
from this we may deduce that particle-hole mode as,
\begin{equation}
 {\tilde{\omega}}_{i}( {\bf{q}} ) = \omega_{ {\bf{k}}_{i} }({\bf{q}})
 + (\frac{ v_{ {\bf{q}} } }{V})\Lambda^{2}_{ {\bf{k}}_{i} }(-{\bf{q}})
\end{equation}
 as is clear from the above derivation, two points must be borne in mind,
 one is, we have to defer the taking of the thermodynamic limit until
 the very end, the other is to exploit the property of the object 
 $ \Lambda_{ {\bf{k}} }({\bf{q}}) $, namely, if 
 $ |{\bf{q}}| << k_{f} $, and $ \Lambda_{ {\bf{k}} }(-{\bf{q}}) = 1 $ 
($ \Lambda_{ {\bf{k}} }(-{\bf{q}}) = 0,1\mbox{  }always $)
  then $ |{\bf{k}}| \approx k_{f} $. Alternatively, we can solve
 for $ {\tilde{\omega}}_{i}( {\bf{q}} ) $ as shown below,
\begin{equation}
 {\tilde{\omega}}_{i}( {\bf{q}} ) = \omega_{ {\bf{k}}_{i} }({\bf{q}})
 + (\frac{ v_{ {\bf{q}} } }{V})
\frac{ \Lambda^{2}_{ {\bf{k}}_{i} }(-{\bf{q}}) }
{ 1 + \frac{ v_{ {\bf{q}} } }{V}
\sum_{ {\bf{k}} }\frac{ \Lambda^{2}_{ {\bf{k}} }({\bf{q}}) }
{ \omega_{ {\bf{k}}_{i} }({\bf{q}}) + \omega_{ {\bf{k}} }(-{\bf{q}}) }
- \frac{ v_{ {\bf{q}} } }{V}
\sum_{ {\bf{k}} \neq {\bf{k}}_{i}}\frac{ \Lambda^{2}_{ {\bf{k}} }(-{\bf{q}}) }
{ \omega_{ {\bf{k}}_{i} }({\bf{q}}) - \omega_{ {\bf{k}} }({\bf{q}}) } }
\end{equation}
 We shall find these formulas useful later on when we try to write
 down the propagator. The collective mode in 1D and 3D may be written down as
 shown below,
\begin{equation}
\omega_{c-1D}(q) = (\frac{ |q| }{m})
\sqrt{ \frac{ (k_{f} + q/2)^{2}  - (k_{f} - q/2)^{2}exp(-\lambda(q)) }
{ 1 - exp(-\lambda(q)) } }
\end{equation}
\begin{equation}
\lambda(q) = (\frac{2 \pi q}{m})(\frac{1}{v_{q}})
\end{equation}
we may also write,
\begin{equation}
\omega^{2}_{c-1D}(q) =
 (\frac{ k_{f}q }{m})^{2} + \epsilon_{q}^{2}
+ 2\mbox{ }\epsilon_{q}(\frac{ k_{f}q }{m})coth(\frac{\lambda(q)}{2})
\label{1DDISP}
\end{equation}
In 3D it is more familiar \cite{Mahan}(only for coulomb repulsion),
\begin{equation}
\omega_{c-3D}( {\bf{q}} ) = \omega_{p}[ 1 + \frac{3}{10}
\frac{ (q\mbox{ }v_{f})^{2} }{\omega^{2}_{p}} ]
\label{3DDISP}
\end{equation}
For more general forms of interaction in 3D the answer may be
 obtained by computing the roots of the equation below,
\begin{equation}
1 - \frac{ n_{0} (v_{ {\bf{q}} }q^{2})/m }{\omega^{2}}
\{1 + \frac{1}{ \omega^{2}}[\frac{3}{5}(qv_{F})^{2} - \epsilon^{2}_{ {\bf{q}} }]
\} = 0
\end{equation}
In 2D, the answer is not available in the books and may be deduced after 
 some algebra as,
\begin{equation}
\omega_{c-2D}( {\bf{q}} ) = \frac{ (\frac{k_{f}|{\bf{q}}|}{m})
(1 + \frac{2\pi/m}{v_{ {\bf{q}} } }) }
{ \sqrt{ \frac{4\pi/m}{ v_{ {\bf{q}} } } 
+ (\frac{2\pi/m}{ v_{ {\bf{q}} } })^{2} } }
\end{equation}
After all this, it is relatively simple to deduce the full propagator.
For reference the free propagator is
\[
\langle \psi^{\dagger}({\bf{x}},t)\psi({\bf{x}}^{'},t^{'})\rangle
 = \rho_{0}
e^{-\sum_{ {\bf{k}}, {\bf{q}} \neq 0 }g^{*}_{ {\bf{k}}, {\bf{q}} }({\bf{x}})
g_{ {\bf{k}}, {\bf{q}} }({\bf{x}}) }
\]
\begin{equation}
\times
e^{\sum_{ {\bf{k}}, {\bf{q}} \neq 0 }g^{*}_{ {\bf{k}}, {\bf{q}} }({\bf{x}})
g_{ {\bf{k}}, {\bf{q}} }({\bf{x}}^{'}) e^{i\mbox{ }\omega_{ {\bf{k}} }({\bf{q}})
(t^{'}-t)} }
\end{equation}
\[
\langle \psi({\bf{x}}^{'},t^{'})\psi^{\dagger}({\bf{x}},t)\rangle
 = \rho_{0}
e^{-\sum_{ {\bf{k}}, {\bf{q}} \neq 0 }f^{*}_{ {\bf{k}}, {\bf{q}} }({\bf{x}}^{'})
f_{ {\bf{k}}, {\bf{q}} }({\bf{x}}^{'}) }
\]
\begin{equation}
\times
e^{\sum_{ {\bf{k}}, {\bf{q}} \neq 0 }f^{*}_{ {\bf{k}}, {\bf{q}} }({\bf{x}})
f_{ {\bf{k}}, {\bf{q}} }({\bf{x}}^{'})
e^{i\mbox{ }\omega_{ {\bf{k}} }({\bf{q}})
(t-t^{'})} }
\end{equation}
\begin{equation}
f_{ {\bf{k}}, {\bf{q}} }({\bf{x}})
= e^{i {\bf{q.x}} }(\frac{1}{2 \mbox{ }N\mbox{ }\epsilon_{ {\bf{q}} } })
\Lambda_{ {\bf{k}} }(-{\bf{q}})\omega_{ {\bf{k}} }({\bf{q}})
 + i\mbox{ }U_{ -{\bf{q}} }({\bf{x}})\Lambda_{ {\bf{k}} }(-{\bf{q}})
\end{equation}
\begin{equation}
g_{ {\bf{k}}, {\bf{q}} }({\bf{x}})
= -e^{-i {\bf{q.x}} }(\frac{1}{2 \mbox{ }N\mbox{ }\epsilon_{ {\bf{q}} } })
\Lambda_{ {\bf{k}} }(-{\bf{q}})\omega_{ {\bf{k}} }({\bf{q}})
 + i\mbox{ }U_{ {\bf{q}} }({\bf{x}})\Lambda_{ {\bf{k}} }(-{\bf{q}})
= -f^{*}_{ {\bf{k}}, {\bf{q}} }({\bf{x}})
\end{equation}
and,
\begin{equation}
{\mathcal{Z}}_{0} = e^{i\mbox{ }\sum_{ {\bf{k}}, {\bf{q}} \neq 0 }
U_{0}({\bf{q}})(\frac{1}{2 \mbox{ }N\epsilon_{ {\bf{q}} }})
(\Lambda_{ {\bf{k}} }(-{\bf{q}}))^{2}\omega_{ {\bf{k}} }({\bf{q}}) }
 e^{\frac{1}{2}
\mbox{ }\sum_{ {\bf{k}}, {\bf{q}} \neq 0 }
(\frac{1}{2 \mbox{ }N\epsilon_{ {\bf{q}} }})^{2}
(\Lambda_{ {\bf{k}} }(-{\bf{q}}))^{2}(\omega_{ {\bf{k}} }({\bf{q}}))^{2} }
e^{\frac{1}{2}
\mbox{ }\sum_{ {\bf{k}}, {\bf{q}} \neq 0 }
(U_{0}({\bf{q}}))^{2}
(\Lambda_{ {\bf{k}} }(-{\bf{q}}))^{2} }
\end{equation}
The time evolved field operator in the interacting case is,
 \begin{equation}
 \psi^{\dagger}({\bf{x}},t) =
exp( \sum_{ {\bf{k}}, {\bf{q}}\neq 0, i }U^{i}_{ {\bf{k}}, {\bf{q}} }
({\bf{x}})b^{\dagger}_{i}({\bf{q}})
e^{i\mbox{ }{\tilde{\omega}}_{i}({\bf{q}}) t} )
exp( -\sum_{ {\bf{k}}, {\bf{q}}\neq 0, i }U^{*i}_{ {\bf{k}}, {\bf{q}} }
({\bf{x}})b_{i}({\bf{q}})
e^{-i\mbox{ }{\tilde{\omega}}_{i}({\bf{q}}) t} )
{\mathcal{R}}_{0}{\mathcal{Z}}^{*}_{0}\sqrt{\rho_{0}}
\end{equation}
where,
\begin{equation}
U^{i}_{ {\bf{k}}, {\bf{q}} } =
f^{*}_{ {\bf{k}}, {\bf{q}} }({\bf{x}})
[a_{ {\bf{k}} }({\bf{q}}), b^{\dagger}_{i}({\bf{q}})]
+ f_{ {\bf{k}}, -{\bf{q}} }({\bf{x}})
[a_{ {\bf{k}} }(-{\bf{q}}), b_{i}({\bf{q}})]
\end{equation}
\[
{\mathcal{R}}_{0} =
exp(-\sum_{ {\bf{k}}, {\bf{q}}, i }
f^{*}_{ {\bf{k}}, {\bf{q}} }({\bf{x}})f_{ {\bf{k}}, {\bf{q}} }({\bf{x}})
[b_{i}({\bf{q}}), a^{\dagger}_{ {\bf{k}} }({\bf{q}})]
[a_{ {\bf{k}} }({\bf{q}}), b^{\dagger}_{i}({\bf{q}})])
\]
\[
\times
exp(-\frac{1}{2}\sum_{ {\bf{k}}, {\bf{q}}, i }
 f^{*}_{ {\bf{k}}, {\bf{q}} }({\bf{x}})f^{*}_{ {\bf{k}}, -{\bf{q}} }({\bf{x}})
[a_{ {\bf{k}} }(-{\bf{q}}), b_{i}({\bf{q}})]
[a_{ {\bf{k}} }({\bf{q}}), b^{\dagger}_{i}({\bf{q}})])
\]
\begin{equation}
\times
exp(-\frac{1}{2}\sum_{ {\bf{k}}, {\bf{q}}, i }
 f_{ {\bf{k}}, {\bf{q}} }({\bf{x}})f_{ {\bf{k}}, -{\bf{q}} }({\bf{x}})
[a_{ {\bf{k}} }(-{\bf{q}}), b_{i}({\bf{q}})]
[a_{ {\bf{k}} }({\bf{q}}), b^{\dagger}_{i}({\bf{q}})])
\end{equation}
The two full fermi propagators may be written down as,
\begin{equation}
\langle \psi^{\dagger}({\bf{x}},t)\psi({\bf{x}}^{'},t^{'})\rangle
 = |{\mathcal{R}}_{0}|^{2}|{\mathcal{Z}}_{0}|^{2}\rho_{0}
e^{\sum_{ {\bf{k}}, {\bf{q}}, i }
U^{*i}_{ {\bf{k}}, {\bf{q}} }({\bf{x}})
U^{i}_{ {\bf{k}}, {\bf{q}} }({\bf{x}}^{'})
e^{i\mbox{ }{\tilde{\omega}}_{i}({\bf{q}})(t^{'}-t)}}
\end{equation}
\begin{equation}
\langle \psi({\bf{x}}^{'},t^{'})\psi^{\dagger}({\bf{x}},t)\rangle
 = |{\mathcal{R}}_{0}|^{2}|{\mathcal{Z}}_{0}|^{2}\rho_{0}
e^{\sum_{ {\bf{k}}, {\bf{q}}, i }
U^{*i}_{ {\bf{k}}, {\bf{q}} }({\bf{x}}^{'})
U^{i}_{ {\bf{k}}, {\bf{q}} }({\bf{x}})
e^{i\mbox{ }{\tilde{\omega}}_{i}({\bf{q}})(t-t^{'})}}
\end{equation}
 Again, it is desirable to use the trick we used in the bose case,
 namely multiply and divide by the free propagator and in the division
 use the form predicted by the bosonized theory and in the multiplication,
 use the form predicted by elementary considerations. This procedure
 also ensures that inspite of the fact we have not verified 
 that the fermi fields written down in terms of the bose fields 
 anticommute, the anticommutation rules are forced on the propagators
 by the free propagators which we know anticommute in the right fashion.
 This leads to the following forms for the propagators,
\begin{equation}
\langle \psi^{\dagger}({\bf{x}},t)\psi({\bf{x}}^{'},t^{'})\rangle
 = |{\mathcal{R}}_{0}|^{2}|{\mathcal{Z}}_{0}|^{4}
e^{\sum_{ {\bf{k}}, {\bf{q}}, i }
U^{*i}_{ {\bf{k}}, {\bf{q}} }({\bf{x}})
U^{i}_{ {\bf{k}}, {\bf{q}} }({\bf{x}}^{'})
e^{i\mbox{ }{\tilde{\omega}}_{i}({\bf{q}})(t^{'}-t)}}
e^{-\sum_{ {\bf{k}}, {\bf{q}} }
g^{*}_{ {\bf{k}}, {\bf{q}} }({\bf{x}})
g_{ {\bf{k}}, {\bf{q}} }({\bf{x}}^{'})
e^{i\mbox{ }\omega_{ {\bf{k}} }({\bf{q}})(t^{'}-t)}}
\langle \psi^{\dagger}({\bf{x}},t)\psi({\bf{x}}^{'},t^{'})\rangle_{free}
\end{equation}
\begin{equation}
\langle \psi({\bf{x}}^{'},t^{'})\psi^{\dagger}({\bf{x}},t)\rangle
 = |{\mathcal{R}}_{0}|^{2}|{\mathcal{Z}}_{0}|^{4}
e^{\sum_{ {\bf{k}}, {\bf{q}}, i }
U^{*i}_{ {\bf{k}}, {\bf{q}} }({\bf{x}}^{'})
U^{i}_{ {\bf{k}}, {\bf{q}} }({\bf{x}})
e^{i\mbox{ }{\tilde{\omega}}_{i}({\bf{q}})(t-t^{'})}}
e^{-\sum_{ {\bf{k}}, {\bf{q}} }
f^{*}_{ {\bf{k}}, {\bf{q}} }({\bf{x}})
f_{ {\bf{k}}, {\bf{q}} }({\bf{x}}^{'})
e^{i\mbox{ }\omega_{ {\bf{k}} }({\bf{q}})(t-t^{'})}}
\langle \psi({\bf{x}}^{'},t^{'})\psi^{\dagger}({\bf{x}},t)\rangle_{free}
\end{equation}
 In the above formula, the index $ i $ runs over both the collective
 mode as well as the particle-hole mode($ i =  c, {\bf{k}}_{i} $)
 The momentum distribution may be evaluated in a different way by computing
 the expectation value of the number operator in Eq.(~\ref{NUMBER}). This leads
 to the following answer. It includes contribution from both particle-hole
 mode and the collective mode. In Appendix D, we show how to derive the same
 momentum distribution using the equation of motion approach(actually just the
 collective part, for purposes of illustration). The full momentum 
 distribution including the particle-hole mode is given below.
\begin{equation}
\langle c^{\dagger}_{ {\bf{k}} }c_{ {\bf{k}} } \rangle
 = \theta(k_{f} - |{\bf{k}}|)F_{1}({\bf{k}})
 + (1 - \theta(k_{f} - |{\bf{k}}|))F_{2}({\bf{k}}) 
\end{equation}
\begin{equation}
F_{1}({\bf{k}}) = 1 - \sum_{ i, {\bf{q}} }
\frac{ 1 - n_{F}({\bf{k}} + {\bf{q}}) }{({\tilde{\omega}}_{i}(-{\bf{q}}) 
+ \frac{ {\bf{k.q}} }{m} + \epsilon_{ {\bf{q}} })^{2} }
g^{2}_{i}(-{\bf{q}})
\end{equation}
\begin{equation}
F_{2}({\bf{k}}) = \sum_{ i, {\bf{q}} }
\frac{ n_{F}({\bf{k}} - {\bf{q}}) }{({\tilde{\omega}}_{i}(-{\bf{q}})
+ \frac{ {\bf{k.q}} }{m} - \epsilon_{ {\bf{q}} })^{2} }
g^{2}_{i}(-{\bf{q}})
\end{equation}
 In the above sum over $ i $, one must include both the collective
 mode and the particle-hole mode($ i = c, {\bf{k}}_{i} $).
 A more general result is possible for systems that are significantly
 more nonideal. This comes about when one does not use the zero-temperature
 non-interacting values in the fermi-bilinear sea-boson correspondence.
 The form of the momentum distribution suggested by this is given
 in Appendix D. 
 It is now very easy to write down a criterion for the breakdown
 of fermi-liquid behaviour. It is given by equating the step at the 
 fermi surface to zero(the quasi-particle residue).
\begin{equation}
Z_{f} = F_{1}(k_{f}) - F_{2}(k_{f}) = 0
\end{equation}
 In the end, it is pertinent to address the claim made in the abstract
 namely that we are able to capture short-wavelength behaviour. 
 The real issue here is that we have two length scales, one is
 the inverse of the fermi momentum the other is the lattice spacing.
 When one speaks of short wavelengths, one means wavelengths comparable to
 the lattice spacing. In the ultra-high density limit, where all the answers 
 we have been deriving are valid, the inverse of the fermi momentum is 
 much too small (compared to the lattice spacing) for the wavelength of
 any external field to be comparable to it. In other words, even if you
 have an external field that varies so rapidly in space that it changes sign
 every lattice spacing, the effective field induced by such an external field
 is still described by the RPA. To put it yet another way, the
  RPA is exact in the 
 ultra-high density limit. Some have argued that this limit is uninteresting
 since in this limit, the coulomb interaction is completely screened out
 and therefore in this regime we just have a fermi liquid. We find that 
 this argument is not entirely true. In fact, we have shown\cite{Setlur}
 that when the inverse of the 
 fermi momentum is small compared to the lattice spacing, it is still
 possible to increase the value of the dimensionless coupling strength
 sufficiently so that fermi liquid behaviour is destroyed. 
 Actually, our results turn upside down almost all conventional
 wisdom, first we find that fermi liquid behaviour persists in 1D 
 for sufficiently weak coupling strengths (even when we assume
 the interactions are hard core delta-function interactions),
 in contrast to the Lieb-Mattis solution of the Tomonaga-Luttinger model.
 Second, we find that fermi-liquid behaviour breaks down in more than
 one dimension for sufficiently strong values of the coupling strength
 in contrast to the answers obtained by Castro-Neto and Fradkin \cite{Neto}.
 In fact we find that fermi liquid behaviour persists in all three dimensions
 for sufficiently small values of the coupling strength and is destroyed
 in all three dimensions for sufficiently large values of the coupling
 strength. 
 It may be argued by the reader that our results are not foolproof either,
 for one, we have neglected several terms in the hamiltonian
 and those terms are small only in the limit when RPA is exact.
 The other points are the technical
 shortcomings, such as the fact that we have not proved the fermi case
 as rigorously as the bose case, like the fermi commutation rules 
 are not explictly verified e.t.c.. Notwithstanding all these shortcomings,
 a strong case is to be made for the revision of entrenched dogma about
 fermi and luttinger liquids.

\section{Conclusions}

 Let us summarise the results obtained so far. We have succeeded in reducing
 to quadratures the propagators of both bose and fermi systems. We have also
 computed the momentum distribution of interacting fermi systems and written
 down a formula for the quasi-particle residue in terms of the
 electron-electron repulsion.
 From this we obtain a criterion for the breakdown of fermi-liquid behaviour.
 The results we obtain contradict widely held views about 1D systems,
 in particular the Lieb-Mattis solution\cite{Lieb}
 of the Tomonaga-Luttinger model
 suggests that the momentum distribution of a 1D system with delta-function
 interactions exhibits no
 discontinuity at the fermi momentum. This is in contrast with the exact
 formula above that does in fact exhibit such a discontinuity for sufficiently
 weak values of the coupling strength and is destroyed
 only for larger values of the coupling strength.
 We attribute this
 discrepency to flaws in the linearised dispersion model (i.e.
 Tomonaga-Luttinger model) and the ensuing
 algebraic manipulations, not all of it transparent. In any event,
 the authors feel that the Tomonaga-Luttinger
 model is not a caricature of the 
 homogeneous interacting fermi system in the ultra-high density limit. 
 Luttinger Liquid theory is based on the unproven
 assumption that the low-energy behaviour of the homogeneous
 interacting fermi system in one-dimension
 is correctly described by the exactly
 solvable Tomonaga-Luttinger model. Our results show that this is false.
 The important qualitative features of the homogeneous interacting
 fermi system namely the presence or absence of a fermi surface
 cannot be surmised by examining the properties of
 the Tomonaga-Luttinger model, especially when the interactions are weak.

\begin{center} APPENDIX A \end{center}

 In this appendix we prove some assersions made earlier.
 First the definition of
 the condensate displacement annhilation operator.
\begin{equation}
d_{ {\bf{q}}/2 }({\bf{q}}) = (\frac{1}{ \sqrt{N_{0}} })
b^{\dagger}_{ {\bf{0}} }b_{ {\bf{q}} }
\label{DEFN}
\end{equation}
 In order to define the quantity
 $ {\mathcal{O}} = (\frac{1}{ \sqrt{N_{0}} }) $ in a manner acceptable
 to most physicists, we proceed as follows. $ {\mathcal{O}} $ is defined to be
 that operator that commutes with the number operator $ N_{0} $
\begin{equation}
[{\mathcal{O}}, N_{0}] = 0
\end{equation}
 in the basis in which $ N_{0} $ is diagonal and possesses non-zero eigenvalues
 (not an unreasonable assumption considering the fact that even in the most
 strongly interacting systems $ N_{0} $ is macroscopic),
 call them $ \{ N_{0}^{r} \} $, then the matrix elements of $ {\mathcal{O}} $
 in the same basis are going to be $ 1/\sqrt{ N_{0}^{r} } $. Having thus
 provided all the matrix elements, the definition of $ {\mathcal{O}} $
 is complete. We have to now show that $ d_{ {\bf{q}}/2 } ({\bf{q}}) $
 satisfies canonical bose commutation rules.
 The simplest way of doing this is to use the polar decomposition
 of $ b_{ {\bf{0}} } $
\begin{equation}
b_{ {\bf{0}} } = exp(-i\mbox{ }X_{0})\sqrt{N_{0}}
\label{POLAR}
\end{equation}
 here $ X_{0} $ is the hermitian operator canonically conjugate to
 $  N_{0} = b^{\dagger}_{ {\bf{0}} }b_{ {\bf{0}} } $, that is,
 $ [X_{0}, N_{0}] = i $.
 This decomposition correctly reproduces the bose commutation rules
 of $ b_{ {\bf{0}} } $ and $ b^{\dagger}_{ {\bf{0}} } $. For example,
\begin{equation}
[b_{ {\bf{0}} }, b^{\dagger}_{ {\bf{0}} }] =
b_{ {\bf{0}} }b^{\dagger}_{ {\bf{0}} } - b^{\dagger}_{ {\bf{0}} }b_{ {\bf{0}} }
 = exp(-i\mbox{ }X_{0})\mbox{  }N_{0}\mbox{  }exp(i\mbox{ }X_{0})
 - N_{0}
 = 1
\end{equation}
 This means that $ d_{ {\bf{q}}/2 }({\bf{q}}) = z^{*}_{0}b_{ {\bf{q}} } $,
 where  $ z^{*}_{0} = exp(i\mbox{ }X_{0}) $. Since,
 $ [z_{0}, b_{ {\bf{q}} }] = 0 $ and $ [z_{0}, b^{\dagger}_{ {\bf{q}} }] = 0 $,
 and $ [z_{0}, z_{0}^{*}] = 0 $, it follows that $ d_{ {\bf{q}}/2 }({\bf{q}}) $
 and $ b_{ {\bf{q}} } $ both satisfy the same commutation rules since,
 $ z^{*}_{0} $ now behaves effectively as a c-number(as regards commutation
 rules with $ b_{ {\bf{q}} } $, $ b^{\dagger}_{ {\bf{q}} } $and $ z_{0} $.
 It is worthwhile pointing out this fact,
\[
[d_{ {\bf{q}}/2 }({\bf{q}}), N_{0}] \neq 0
\]
rather,
\begin{equation}
[d_{ {\bf{q}}/2 }({\bf{q}}), N] = 0
\end{equation}
though not obviously so.
In order to prove this,
\[
[d_{ {\bf{q}}/2 }({\bf{q}}),N] = [d_{ {\bf{q}}/2 }({\bf{q}}),N_{0}]
 + [d_{ {\bf{q}}/2 }({\bf{q}}),\sum_{ {\bf{q}}^{'} \neq 0 }
b^{\dagger}_{ {\bf{q}}^{'} }b_{ {\bf{q}}^{'} }]
\]
\[
 = [exp(i\mbox{  }X_{0}),N_{0}] b_{ {\bf{q}} }
+ exp(i\mbox{  }X_{0})\sum_{ {\bf{q}}^{'} \neq 0 }
 [b_{ {\bf{q}} },b^{\dagger}_{ {\bf{q}}^{'} }b_{ {\bf{q}}^{'} }]
\]
\[
 = ( exp(i\mbox{  }X_{0})N_{0} - N_{0}exp(i\mbox{  }X_{0}) )b_{ {\bf{q}} }
+ exp(i\mbox{  }X_{0})b_{ {\bf{q}}  }
\]
\begin{equation}
 = [i\mbox{  }X_{0}, N_{0}]exp(i\mbox{  }X_{0}) b_{ {\bf{q}} }
+ exp(i\mbox{  }X_{0})b_{ {\bf{q}}  } = -exp(i\mbox{  }X_{0}) b_{ {\bf{q}} }
+ exp(i\mbox{  }X_{0})b_{ {\bf{q}}  } = 0
\end{equation}
 Next, one would like to prove the equation Eq.(~\ref{BOSE}). For this we
 simply plug in the definition  Eq.(~\ref{DEFN}) into Eq.(~\ref{BOSE})
 and verify that is reduces to an identity. The details are as follows,
 first define,
\[
L_{ {\bf{k}}, {\bf{q}} } = N_{0}
 \delta_{ {\bf{k}}, 0 }\delta_{ {\bf{q}}, 0 } +
[\delta_{ {\bf{k+q/2}}, 0 }(\sqrt{N_{0}})d_{ {\bf{k}} }(-{\bf{q}})
 + \delta_{ {\bf{k-q/2}}, 0 }d^{\dagger}_{ {\bf{k}} }({\bf{q}})
(\sqrt{N_{0}})]
\]
\begin{equation}
+
d^{\dagger}_{ (1/2)({\bf{k+q/2}}) }({\bf{k+q/2}})
d_{ (1/2)({\bf{k-q/2}}) }({\bf{k-q/2}})
\end{equation}
The proof involves these cases,
\newline
(i) $ {\bf{k}} = 0 $ and $ {\bf{q}} = 0 $
\newline
In this case,
\begin{equation}
L_{ {\bf{0}}, {\bf{0}} } = N_{0} = b^{\dagger}_{0}b_{0}
\end{equation}
\newline
(ii)  $ {\bf{k}} + {\bf{q}}/2 = 0 $ but $ {\bf{k}} - {\bf{q}}/2 \neq 0 $
\newline
\begin{equation}
L_{ {\bf{k}} = -{\bf{q}}/2, {\bf{q}} } =
 (\sqrt{N_{0}})d_{ -{\bf{q}}/2 }(-{\bf{q}})
 = b^{\dagger}_{0}b_{ -{\bf{q}} }
\end{equation}
\newline
(iii)  $ {\bf{k}} - {\bf{q}}/2 = 0 $ but $ {\bf{k}} + {\bf{q}}/2 \neq 0 $
\newline
\begin{equation}
L_{ {\bf{k}} = {\bf{q}}/2, {\bf{q}} } =
 d^{\dagger}_{ {\bf{q}}/2 }({\bf{q}})(\sqrt{N_{0}})
 = b^{\dagger}_{ {\bf{q}} }b_{ {\bf{0}} }
\end{equation}
\newline
(iii)  $ {\bf{k}} - {\bf{q}}/2 \neq 0 $ and $ {\bf{k}} + {\bf{q}}/2 \neq 0 $
\newline
\begin{equation}
L_{ {\bf{k}}, {\bf{q}} } =
 d^{\dagger}_{ (1/2)({\bf{k}} + {\bf{q}}/2) }({\bf{k}} + {\bf{q}}/2)
d_{ (1/2)({\bf{k}} - {\bf{q}}/2) }({\bf{k}} - {\bf{q}}/2)
 = b^{\dagger}_{ {\bf{k}} + {\bf{q}}/2 }exp(-i\mbox{ }X_{0})
exp(i\mbox{ }X_{0})b_{ {\bf{k}} - {\bf{q}}/2 }
 = b^{\dagger}_{ {\bf{k}} + {\bf{q}}/2 }b_{ {\bf{k}} - {\bf{q}}/2 }
\end{equation}
Therefore in all cases,
\begin{equation}
 L_{ {\bf{k}}, {\bf{q}} } =
b^{\dagger}_{ {\bf{k}}+{\bf{q}}/2 }b_{ {\bf{k}}-{\bf{q}}/2 }
\end{equation}
 and thus Eq.(~\ref{BOSE}) follows.  Finally, we would like to clarify the
 finite temperature case. In particular, what is the chemical potential
 of the condensate displacement bosons ? Is it zero or is it the same as
 that of the parent bosons ?  The answer may be found by computing the
 thermodynamic expectation value of the number of bosons in the condensate
 $ N_{0} $,
\begin{equation}
\langle N_{0} \rangle = N - \sum_{ {\bf{q}} \neq 0 }\langle
 d^{\dagger}_{ (1/2){\bf{q}} }({\bf{q}}) d_{ (1/2){\bf{q}} }({\bf{q}})
\rangle
\end{equation}
We also know the answer from elementary considerations, it is,
\begin{equation}
\langle N_{0} \rangle = N - \sum_{ {\bf{q}} \neq 0 }
 \frac{1}{exp(\beta(\epsilon_{ {\bf{q}} }-\mu)) -  1}
\end{equation}
where $ \mu $ is the chemical potential of the parent bosons. Then it follows
 that,
\begin{equation}
\langle d^{\dagger}_{ (1/2){\bf{q}} }({\bf{q}}) d_{ (1/2){\bf{q}} }({\bf{q}})
\rangle =  \frac{1}{exp(\beta(\epsilon_{ {\bf{q}} }-\mu)) -  1}
\end{equation}
 In other words, the chemical potential of the condensate displacement bosons
 is the same as that of the parent bosons.
\begin{equation}
\mu_{parent} = \mu_{cond/displ}
\end{equation}


\begin{center} APPENDIX B \end{center}

 In this Appendix, we try to to make plausible the correspondence between the
 number-conserving product of two fermi fields and the sea-bosons.
 Let us rewrite the bose case (Eq.(~\ref{BOSE})) more suggestively,
\[
b^{\dagger}_{ {\bf{k+q/2}} }b_{ {\bf{k-q/2}} }
 =  O({\bf{k}})\mbox{  }\delta_{ {\bf{q}}, {\bf{0}} }
 +
[\sqrt{n_{ {\bf{k}} + {\bf{q}}/2 }}A_{ {\bf{k}} }(-{\bf{q}})
 + A^{\dagger}_{ {\bf{k}} }({\bf{q}})\sqrt{n_{ {\bf{k}} - {\bf{q}}/2 }}]
\]
\[
 +
 \sum_{ {\bf{q}}_{1} }
  A^{\dagger}_{ {\bf{k}} + {\bf{q}}/2 - {\bf{q}}_{1}/2 }({\bf{q}}_{1})
A_{ {\bf{k}} -  {\bf{q}}_{1}/2 }(-{\bf{q}} + {\bf{q}}_{1})
\]
\begin{equation}
 - \sum_{ {\bf{q}}_{1} }
  A^{\dagger}_{ {\bf{k}} - {\bf{q}}/2 + {\bf{q}}_{1}/2 }({\bf{q}}_{1})
A_{ {\bf{k}} + {\bf{q}}_{1}/2 }(-{\bf{q}} + {\bf{q}}_{1})
\label{SUGG}
\end{equation}
In the bose case, 
\begin{equation}
 A_{ {\bf{k}} }({\bf{q}}) = \delta_{ {\bf{k}}-{\bf{q}}/2, 0}
d_{ {\bf{q}}/2 }({\bf{q}})
\end{equation}
and,
\begin{equation}
O({\bf{k}}) = N\mbox{  }\delta_{ {\bf{k}}, 0 }
\end{equation}
 Observe that the suggestively extravagant notation in Eq.(~\ref{SUGG})
 is meant to imply that a very similar relation holds
 in the fermi case which we reproduce below,
\[
c^{\dagger}_{ {\bf{k+q/2}} }c_{ {\bf{k-q/2}} }
 =  O({\bf{k}})\mbox{  }\delta_{ {\bf{q}}, {\bf{0}} }
 +
[\sqrt{n_{ {\bf{k}} + {\bf{q}}/2 }}A_{ {\bf{k}} }(-{\bf{q}})
 + A^{\dagger}_{ {\bf{k}} }({\bf{q}})\sqrt{n_{ {\bf{k}} - {\bf{q}}/2 }}]
\]
\[
 + \sum_{ {\bf{q}}_{1} }
  A^{\dagger}_{ {\bf{k}} + {\bf{q}}/2 - {\bf{q}}_{1}/2 }({\bf{q}}_{1})
A_{ {\bf{k}} -  {\bf{q}}_{1}/2 }(-{\bf{q}} + {\bf{q}}_{1})
\]
\begin{equation}
 -  \sum_{ {\bf{q}}_{1} }
  A^{\dagger}_{ {\bf{k}} - {\bf{q}}/2 + {\bf{q}}_{1}/2 }({\bf{q}}_{1})
A_{ {\bf{k}} + {\bf{q}}_{1}/2 }(-{\bf{q}} + {\bf{q}}_{1})
\label{FERMI}
\end{equation} 
 Here $ A_{ {\bf{k}} }({\bf{q}}) $ depends on two momentum labels unlike in
 the bose case. This has to do with the fact the now $ O({\bf{k}}) $ no longer
 has the simple structure we saw in the bose case. We must now invert this
 relation and obtain a formula for the operator $ A_{ {\bf{k}} }({\bf{q}}) $.
 It is not at all clear that this object will behave like an exact boson
 annhilation operator. The alternative is to write down an ansatz for an 
 exact boson in analogy with the bose case and determine the unknown in the
 formula by imposing canonical bose commutation rules. 
\begin{equation}
a_{ {\bf{k}} }({\bf{q}}) = \frac{1}{\sqrt{n_{ {\bf{k}} - {\bf{q}}/2 }}}
c^{\dagger}_{ {\bf{k}} - {\bf{q}}/2 }M({\bf{k}}, {\bf{q}})
c_{ {\bf{k}} + {\bf{q}}/2 }
\end{equation}
 The unknown operator $ M({\bf{k}}, {\bf{q}}) $ has to be related to some
 number-conserving fermi bilinear by demanding that the operator
 $ a_{ {\bf{k}} }({\bf{q}}) $ obey canonical bose commutation rules,
\begin{equation}
[a_{ {\bf{k}} }({\bf{q}}), a_{ {\bf{k}}^{'} }({\bf{q}}^{'})] = 0
\end{equation}
\begin{equation}
[a_{ {\bf{k}} }({\bf{q}}), a^{\dagger}_{ {\bf{k}}^{'} }({\bf{q}}^{'})] = 
\delta_{ {\bf{k}}, {\bf{k}}^{'} }\delta_{ {\bf{q}}, {\bf{q}}^{'} }
\end{equation}
 It is at present beyond the authors to arrive at a formula for 
 $ M({\bf{k}},{\bf{q}}) $. Notwithstanding this, it is still useful to capture
 some sort of an approximate correspondence like the one introduced in
 Sec. II.
 The relations written down there have the following positive features
\newline
(i)They recover the RPA dielectric function at zero and finite temperatures.
\newline
(ii)They capture the correct four-point and six-point functions at zero and 
 finite temperatures.
\newline
(iii)The formula for the sea-boson in Eq.(~\ref{SEABOSONFORM}) when
 plugged into the correspondence for the number operator in 
 Eq.(~\ref{NUMBER}) gives an identity.
\newline
The only negative aspect of this correspondence is that 
\newline
(I) The mutual commutation rules between the off-diagonal fermi bilinears
 is recovered correctly 
 only upto terms linear in the sea-bosons. That is, somehow the
 operators on the right side of these commutations rules should not be too 
 different from their approximations obtained by
 dropping terms higher than the linear order.
 This is no doubt a strong assumption. This is 
 in fact equivalent to RPA(perhaps even better than RPA). 
\newline
\newline
 The definition of the seaboson is incomplete without the prescription
 of the phase $ \theta({\bf{k}}, {\bf{q}}) $. In order to derive an
 expression for this, we again make heavy use of the bose case which
 we have proved rigorously in Appendix A. There we found that plugging
 in the expression for the condensate-displacement boson into the 
 correspondence resulted in an identity in case of $ {\bf{q}} \neq 0 $
 (the $  {\bf{q}} = 0 $ case being special). This identity comes about
 in a very specific fashion. In the general form of the correspondence
 outlined in Eq.(~\ref{SUGG}), we find that the sum on the right that comes
 with a negative sign is identically zero( for $ {\bf{q}} \neq 0 $ )
 and the sum on the right that comes with a positive sign is equal
 to the left-hand side, except in "rare" cases when either 
 $ {\bf{k}}+{\bf{q}}/2 = 0 $ or $ {\bf{k}}-{\bf{q}}/2 = 0 $. We shall
 adopt the same approach in the fermi casse 
 and try to fix the phase $ \theta({\bf{k}}, {\bf{q}}) $
 such that the identity is satisfied in the manner just described. 
 Let us now write down the potential identity,
\[
c^{\dagger}_{ {\bf{k}} + {\bf{q}}/2 }c_{ {\bf{k}} - {\bf{q}}/2 }
 = \Lambda_{ {\bf{k}} }({\bf{q}})
\frac{1}{\sqrt{ n_{ {\bf{k}} + {\bf{q}}/2 } }}
c^{\dagger}_{ {\bf{k}} + {\bf{q}}/2 }(\frac{ n^{\beta}({\bf{k}}+{\bf{q}}/2)}
{\langle N \rangle })^{\frac{1}{2}}
e^{i\mbox{    }\theta({\bf{k}},-{\bf{q}})}
c_{ {\bf{k}} - {\bf{q}}/2 }
\]
\[
+ \Lambda_{ {\bf{k}} }(-{\bf{q}})
c^{\dagger}_{ {\bf{k}} + {\bf{q}}/2 }
e^{-i\mbox{    }\theta({\bf{k}},{\bf{q}})}
(\frac{ n^{\beta}({\bf{k}}-{\bf{q}}/2)}
{\langle N \rangle })^{\frac{1}{2}}c_{ {\bf{k}} - {\bf{q}}/2 }
\frac{1}{\sqrt{ n_{ {\bf{k}} - {\bf{q}}/2 } }}
\]
\[
+ T_{1}({\bf{k}}, {\bf{q}})
c^{\dagger}_{ {\bf{k}} + {\bf{q}}/2 }
(\sum_{ {\bf{q}}_{1} \neq {\bf{q}}, {\bf{0}} }
\frac{  n^{\beta}({\bf{k}}+{\bf{q}}/2 - {\bf{q}}_{1})}
{\langle N \rangle }
e^{i\mbox{  }\theta({\bf{k}} - {\bf{q}}_{1}/2, -{\bf{q}}+{\bf{q}}_{1})}
e^{-i\mbox{  }\theta({\bf{k}} + {\bf{q}}/2 - {\bf{q}}_{1}/2, {\bf{q}}_{1})})
c_{ {\bf{k}} - {\bf{q}}/2 }
\]
\[
- T_{2}({\bf{k}}, {\bf{q}})
c_{ {\bf{k}} - {\bf{q}}/2 }\frac{1}{\sqrt{ n_{ {\bf{k}} - {\bf{q}}/2 }}}
\frac{1}{\sqrt{ n_{ {\bf{k}} + {\bf{q}}/2 }}}
c^{\dagger}_{ {\bf{k}} + {\bf{q}}/2 }
(\frac{ n^{\beta}({\bf{k}}+{\bf{q}}/2) }{\langle N \rangle })^{\frac{1}{2}}
(\frac{ n^{\beta}({\bf{k}}-{\bf{q}}/2) }{\langle N \rangle })^{\frac{1}{2}}
\]
\begin{equation}
\times
\sum_{ {\bf{q}}_{1} \neq {\bf{q}}, {\bf{0}} }
n_{ {\bf{k}} - {\bf{q}}/2 + {\bf{q}}_{1} }
e^{i\mbox{  }\theta({\bf{k}} + {\bf{q}}_{1}/2, -{\bf{q}} + {\bf{q}}_{1})}
e^{-i\mbox{  }\theta({\bf{k}} - {\bf{q}}/2  + {\bf{q}}_{1}/2, {\bf{q}}_{1})}
\end{equation}
 Here, since we are not involved in proving the rigorous correspondence, but
 just the salient features, we are entitled to some leeway. In particular,
 we shall turn a blind eye to the fact that there exist these objects
 $ T_{1}({\bf{k}},{\bf{q}}) $ and  $ T_{2}({\bf{k}},{\bf{q}}) $, in
 fact set them both equal to unity, just for the moment. 
 The exact correspondence in terms of the $ A_{ {\bf{k}} }({\bf{q}}) $ 
 seems to suggest exactly this. Then we find that, if we choose our
 $ \theta({\bf{k}},{\bf{q}}) $ to be such that 
\begin{equation}
\theta({\bf{k}} - {\bf{q}}_{1}/2, -{\bf{q}}+{\bf{q}}_{1})
= \theta({\bf{k}} + {\bf{q}}/2 - {\bf{q}}_{1}/2, {\bf{q}}_{1})
\end{equation}
and,
\begin{equation}
\sum_{ {\bf{q}}_{1}\neq {\bf{0}}, {\bf{q}} }
{\bar{n}}_{ {\bf{k}} - {\bf{q}}/2 + {\bf{q}}_{1} }
e^{i\mbox{  }\theta({\bf{k}} + {\bf{q}}_{1}/2, -{\bf{q}} + {\bf{q}}_{1})}
e^{-i\mbox{  }\theta({\bf{k}} - {\bf{q}}/2  + {\bf{q}}_{1}/2, {\bf{q}}_{1})}
 = 0
\end{equation}
 then all is well. Terms that were linear in the sea-bosons are 
 vanishingly small in the thermodynamic limit, and are important only
 when both the sums on the right side are identically zero for some
 reason, that is, it is "rarely" important just like in the bose case.
 It is not really important to write down an explicit formula for the
 phase function $ \theta({\bf{k}}, {\bf{q}}) $, it is merely
 sufficient to show that it does what is required of it, namely, it provides
 the "random phase" that cancels terms that enable the whole machinery to
 run smoothly. Lastly, we have not yet verified that this sea-boson
 obeys canonical commutation rules. This is again a tricky problem, it
 is likely to be resolved by the exact approach which is beyond the scope
 of this article. It is merely sufficient to point out that this is likely to
 come about due to the strong likelyhood that the phase 
 $ \theta({\bf{k}},{\bf{q}}) $ is actually a functional of the number operator.


  The correspondence that we have just defended is nothing but a more elegant
  version of the correspondence introduced by the pioneers like 
  Castro-Neto and Fradkin\cite{Neto}. Any criticism that may be leveled
  against our approach may equally well be leveled against theirs. The only
  difference between our approach and theirs is that the single-particle
  properties which they are so fervently seeking are far more elegantly 
  recovered by our approach since we do not linearise the bare fermion
  dispersion or use the clumsy Luther construction\cite{Luther}.
  Indeed, we have even shown that the answers
  for the 1D case are different from the Tomonaga-Luttinger model that 
  linearise the bare fermion dispersion. 

  The other issue worth addressing at this stage is the validity of
  the prescription in Eq.(~\ref{PRES}). It can be seen from the exact
  correspondence in Eq.(~\ref{FERMI}) that as 
  $ {\bf{q}} \rightarrow {\bf{0}} $ terms that correspond to
  corrections to the RPA-form of the full hamiltonian 
   vanish at least as fast as $ |{\bf{q}}|/k_{f} $.
  The RPA-terms themselves don't vanish and tend toward
  ($ \lim_{ {\bf{q}} \rightarrow {\bf{0}} } A_{ {\bf{k}} }(-{\bf{q}}) \neq 0 $
 as in the bose case) 
\begin{equation}
\sum_{ {\bf{k}} }\mbox{        }
{\sqrt{n_{ {\bf{k}} }}}A_{ {\bf{k}} }(-{\bf{q}})
 + \sum_{ {\bf{k}} }
\mbox{        }A^{\dagger}_{ {\bf{k}} }({\bf{q}}){\sqrt{ n_{ {\bf{k}} } }}
\end{equation}
  In order for the prescription in Eq.(~\ref{PRES}) to be accurate,
  it is important for the interaction $ v_{ {\bf{q}} } $
  to possess these properties, first it must vanish for large enough
 $ {\bf{q}} $(or small inter-particle separation),
\begin{equation}
Lim_{ |{\bf{q}}| \rightarrow c_{0}\mbox{   }k_{f} }\mbox{                  	 }
v_{ {\bf{q}} } \rightarrow 0
\end{equation} 
  where $ c_{0} $ is small compared to unity and positive. This ensures
  that the only possible contributions come from small $ {\bf{q}} $ where
  corrections to the RPA-form themselves are small. In addition, 
   if we also make 
  sure that the interaction vanishes fast enough for large inter-particle
  seperations so that,  
\begin{equation}
Lim_{ |{\bf{q}}| \rightarrow 0 }\mbox{     } v_{ {\bf{q}} }
 \rightarrow |{\bf{q}}|^{D}
\end{equation}
 where $ D = 0,1,2,.. $(larger the better), then our formalism is in fact
 EXACT as $ k_{f} \rightarrow \infty $(or sufficiently large).  It may
 be argued that this state of affairs is most likely uninteresting
 since it may not be realisable in practice, when it is, it merely leads
 to a fermi liquid. This is a valid point. 
 But it is worth pointing out that non-fermi liquid behaviour can still
 emerge in such systems when the interaction strength
 (with the same functional form) becomes strong enough. These considerations
 also tell us that for an interaction of the delta-function type in
 1D, provided the strength is weak enough, we have a fermi-liquid in 
 contrast to the Lieb-Mattis solution of the Tomonaga-Luttinger model.
 
  In any event, the philosophy is, that having introduced
  sea-bosons, we more or less forget about the fact that it was fermions that
  motivated their introduction in the first place
  and instead try to write down a whole new set
  of models in terms of the sea-bosons calibrate them appropriately so
  that they capture the salient features of the real world. It is not
  a tautology to remark that we have in our hands a whole class of exactly
  solvable models of correlated fermions that is easier to use than mean-field
  theory itself but capture effects significantly beyond diagrammatic
  perturbation theory, like the nonanalytic dependence of the
  momentum distribution
  on the coupling strength(written down in Appendix D).

\begin{center} APPENDIX C \end{center}

 In this appendix we demonstrate that the RPA dielectric function
 is recovered exactly by selectively
 retaining parts of the coulomb interaction that lead to RPA.
 We know that the kinetic energy in the bose language is given by,
\begin{equation}
H_{kin} = \sum_{ {\bf{k}}, {\bf{q}} }
(\frac{ {\bf{k.q}} }{m})
a^{\dagger}_{ {\bf{k}} }({\bf{q}})a_{ {\bf{k}} }({\bf{q}})
\end{equation}
 For this let us choose,
\begin{equation}
H_{I} = \sum_{ {\bf{q}} \neq 0 }\frac{ v_{ {\bf{q}} } }{2V}
{\tilde{\rho}}_{ {\bf{q}} }{\tilde{\rho}}_{ -{\bf{q}} }
\end{equation}
 where,
\begin{equation}
{\tilde{\rho}}_{ {\bf{q}} } = \sum_{ {\bf{k}} }
[ \Lambda_{ {\bf{k}} }({\bf{q}})
a_{ {\bf{k}} }(-{\bf{q}})
+ \Lambda_{ {\bf{k}} }(-{\bf{q}})
a^{\dagger}_{ {\bf{k}} }({\bf{q}}) ]
\end{equation}
 From this it may be shown that the RPA dielectric function
 is recovered as the following demonstration shows. Assume that
 a weak time-varying external perturbation is applied as shown below,
\begin{equation}
H_{ext} = \sum_{ {\bf{q}} \neq 0 }
(U_{ext}({\bf{q}},t) + U^{*}_{ext}(-{\bf{q}},t))
{\tilde{\rho}}_{ {\bf{q}} }
\end{equation}
here,
\begin{equation}
 U_{ext}({\vec{r}},t) = U_{0}
\mbox{  }e^{i{\bf{q}}.{\vec{r}} -i \omega \mbox{ }t}
\end{equation}
Let us now write down the equations of motion for the various bose fields,
\[
i\frac{ \partial }{\partial t} \langle a^{t}_{ {\bf{k}} }({\bf{q}}) \rangle
 = \omega_{ {\bf{k}} }({\bf{q}})
\langle a^{t}_{ {\bf{k}} }({\bf{q}})  \rangle
 + (\frac{ v_{ {\bf{q}} } }{V})\Lambda_{ {\bf{k}} }(-{\bf{q}})
\sum_{ {\bf{k}}^{'} }[\Lambda_{ {\bf{k}}^{'} }(-{\bf{q}})
\langle a^{t}_{ {\bf{k}}^{'} }({\bf{q}})  \rangle
+ \Lambda_{ {\bf{k}}^{'} }({\bf{q}})
\langle a^{t\dagger}_{ {\bf{k}}^{'} }(-{\bf{q}})  \rangle ]
\]
\begin{equation}
+ (U_{ext}({\bf{q}},t)+ U^{*}_{ext}(-{\bf{q}},t))
\Lambda_{ {\bf{k}} }(-{\bf{q}})
\end{equation}
\[
-i\frac{ \partial }{\partial t}
\langle a^{t\dagger}_{ {\bf{k}} }(-{\bf{q}}) \rangle
 = \omega_{ {\bf{k}} }(-{\bf{q}})
\langle a^{t\dagger}_{ {\bf{k}} }(-{\bf{q}})  \rangle
 + (\frac{ v_{ {\bf{q}} } }{V})\Lambda_{ {\bf{k}} }({\bf{q}})
\sum_{ {\bf{k}}^{'} }[\Lambda_{ {\bf{k}}^{'} }(-{\bf{q}})
\langle a^{t}_{ {\bf{k}}^{'} }({\bf{q}})  \rangle
+ \Lambda_{ {\bf{k}}^{'} }({\bf{q}})
\langle a^{t\dagger}_{ {\bf{k}}^{'} }(-{\bf{q}})  \rangle ]
\]
\begin{equation}
+ (U_{ext}({\bf{q}},t)+ U^{*}_{ext}(-{\bf{q}},t))
\Lambda_{ {\bf{k}} }({\bf{q}})
\end{equation}
Now, let us decompose the expectation values as follows,
\begin{equation}
\langle a^{t}_{ {\bf{k}} }({\bf{q}})  \rangle
 = U_{ext}({\bf{q}},t)C_{ {\bf{k}} }({\bf{q}})
+ U^{*}_{ext}(-{\bf{q}},t)D_{ {\bf{k}} }({\bf{q}})
\end{equation}
\begin{equation}
\langle a^{t\dagger}_{ {\bf{k}} }(-{\bf{q}})  \rangle
 = U^{*}_{ext}(-{\bf{q}},t)C^{*}_{ {\bf{k}} }(-{\bf{q}})
+ U_{ext}({\bf{q}},t)D^{*}_{ {\bf{k}} }(-{\bf{q}})
\end{equation}
The coefficients $ C_{ {\bf{k}} }({\bf{q}}) $ and
 $ D^{*}_{ {\bf{k}} }(-{\bf{q}}) $ satisfy,
\begin{equation}
\omega \mbox{ }C_{ {\bf{k}} }({\bf{q}}) = \omega_{ {\bf{k}} }({\bf{q}})
C_{ {\bf{k}} }({\bf{q}})
+ (\frac{ v_{ {\bf{q}} } }{V})\Lambda_{ {\bf{k}} }(-{\bf{q}})
\sum_{ {\bf{k}}^{'} }[\Lambda_{ {\bf{k}}^{'} }(-{\bf{q}})
C_{ {\bf{k}}^{'} }({\bf{q}}) + \Lambda_{ {\bf{k}}^{'} }({\bf{q}})
D^{*}_{ {\bf{k}}^{'} }(-{\bf{q}})]
+ \Lambda_{ {\bf{k}} }(-{\bf{q}})
\end{equation}
\begin{equation}
-\omega \mbox{ }D^{*}_{ {\bf{k}} }(-{\bf{q}})
= \omega_{ {\bf{k}} }(-{\bf{q}})
D^{*}_{ {\bf{k}} }(-{\bf{q}})
+ (\frac{ v_{ {\bf{q}} } }{V})\Lambda_{ {\bf{k}} }({\bf{q}})
\sum_{ {\bf{k}}^{'} }[\Lambda_{ {\bf{k}}^{'} }({\bf{q}})
D^{*}_{ {\bf{k}}^{'} }(-{\bf{q}}) + \Lambda_{ {\bf{k}}^{'} }(-{\bf{q}})
C_{ {\bf{k}}^{'} }({\bf{q}})]
+ \Lambda_{ {\bf{k}} }({\bf{q}})
\end{equation}
Now, the effective potential may be written as,
\begin{equation}
U_{eff}({\bf{q}},t) = U_{ext}({\bf{q}},t) +
(\frac{ v_{ {\bf{q}} } }{V})\langle \rho_{ -{\bf{q}} } \rangle^{'}
U_{ext}({\bf{q}},t)
\end{equation}
here,
\begin{equation}
\langle \rho_{ -{\bf{q}} } \rangle
 = U_{ext}({\bf{q}},t)\langle \rho_{ -{\bf{q}} } \rangle^{'}
+ U^{*}_{ext}(-{\bf{q}},t)\langle \rho_{ -{\bf{q}} } \rangle^{''}
\end{equation}
Using the fact that,
\begin{equation}
\langle \rho_{ -{\bf{q}} } \rangle^{'}
 = \sum_{ {\bf{k}} }\Lambda_{ {\bf{k}} }(-{\bf{q}})
C_{ {\bf{k}} }({\bf{q}}) + \sum_{ {\bf{k}} }
\Lambda_{ {\bf{k}} }({\bf{q}})D^{*}_{ {\bf{k}} }(-{\bf{q}})
\end{equation}
 Solving these equations and using the fact that the dielectric function
 is just the ratio of the external divided by the effective
 potential we get,
\begin{equation}
\epsilon( {\bf{q}}, \omega) = \frac{ U_{ext}({\bf{q}},t) }
{ U_{eff}({\bf{q}},t) } 
 = 1 + \frac{ v_{ {\bf{q}} } }{V}\sum_{ {\bf{k}} }
\frac{ n_{F}({\bf{k+q/2}}) - n_{F}({\bf{k-q/2}}) }
{ \omega - \frac{ {\bf{k.q}} }{m} }
\label{EQQ}
\end{equation}
Which is nothing but the RPA dielectric function of Bohm and Pines.
                                
\begin{center} APPENDIX D \end{center}

 In this appendix we use the equation of motion approach to
 solve for the momentum
 distribution and compare it with the solution obtained via exact 
 diagonalisation as described in the main text.
 The equations of motion for the bose propagators read as,
\[
(i\frac{\partial}{\partial t} - \omega_{ {\bf{k}} }({\bf{q}}))
\frac{ -i\langle T a^{t}_{ {\bf{k}} }({\bf{q}})
a^{\dagger}_{ {\bf{k}}^{'} }({\bf{q}}^{'}) \rangle }{\langle T 1 \rangle}
 = \delta_{ {\bf{k}}, {\bf{k}}^{'} }
\delta_{ {\bf{q}}, {\bf{q}}^{'} }\delta(t)
\]
\begin{equation}
+ (\frac{ v_{ {\bf{q}} } }{V})
\Lambda_{ {\bf{k}} }(-{\bf{q}})
\sum_{ {\bf{k}}^{''} }[\Lambda_{ {\bf{k}}^{''} }(-{\bf{q}})
\frac{ -i\langle T a^{t}_{ {\bf{k}}^{''} }({\bf{q}})
a^{\dagger}_{ {\bf{k}}^{'} }({\bf{q}}^{'}) \rangle }{\langle T 1 \rangle}
+ \Lambda_{ {\bf{k}}^{''} }({\bf{q}})
\frac{ -i\langle T a^{\dagger t}_{ {\bf{k}}^{''} }(-{\bf{q}})
a^{\dagger}_{ {\bf{k}}^{'} }({\bf{q}}^{'}) \rangle }{\langle T 1 \rangle}]
\end{equation}
\[
(i\frac{\partial}{\partial t} + \omega_{ {\bf{k}} }(-{\bf{q}}))
\frac{ -i\langle T a^{\dagger t}_{ {\bf{k}} }(-{\bf{q}})
a^{\dagger}_{ {\bf{k}}^{'} }({\bf{q}}^{'}) \rangle }{\langle T 1 \rangle}
\]
\begin{equation}
= -(\frac{ v_{ {\bf{q}} } }{V})
\Lambda_{ {\bf{k}} }({\bf{q}})
\sum_{ {\bf{k}}^{''} }[\Lambda_{ {\bf{k}}^{''} }({\bf{q}})
\frac{ -i\langle T a^{\dagger t}_{ {\bf{k}}^{''} }(-{\bf{q}})
a^{\dagger}_{ {\bf{k}}^{'} }({\bf{q}}^{'}) \rangle }{\langle T 1 \rangle}
+ \Lambda_{ {\bf{k}}^{''} }(-{\bf{q}})
\frac{ -i\langle T a^{t}_{ {\bf{k}}^{''} }({\bf{q}})
a^{\dagger}_{ {\bf{k}}^{'} }({\bf{q}}^{'}) \rangle }{\langle T 1 \rangle} ]
\end{equation}
The boundary conditions on these propagators may be written down as,
\begin{equation}
\frac{ -i\langle T a^{\dagger t}_{ {\bf{k}} }(-{\bf{q}})
a^{\dagger}_{ {\bf{k}}^{'} }({\bf{q}}^{'}) \rangle }{\langle T 1 \rangle}
 = \frac{ -i\langle T a^{\dagger (t - i\beta)}_{ {\bf{k}} }(-{\bf{q}})
a^{\dagger}_{ {\bf{k}}^{'} }({\bf{q}}^{'}) \rangle }{\langle T 1 \rangle}
\end{equation}
\begin{equation}
\frac{ -i\langle T a^{t}_{ {\bf{k}} }({\bf{q}})
a^{\dagger}_{ {\bf{k}}^{'} }({\bf{q}}^{'}) \rangle }{\langle T 1 \rangle}
 = \frac{ -i\langle T a^{(t - i\beta)}_{ {\bf{k}} }({\bf{q}})
a^{\dagger}_{ {\bf{k}}^{'} }({\bf{q}}^{'}) \rangle }{\langle T 1 \rangle}
\end{equation}
\begin{equation}
\delta(t) = (\frac{1}{-i \mbox{ }\beta})\sum_{n} exp(\omega_{n}t)
\end{equation}
\begin{equation}
\theta(t) = (\frac{1}{-i \mbox{ }\beta})\sum_{n}
\frac{ exp(\omega_{n}t) }{\omega_{n}}
\end{equation}
The boundary conditions imply that we may write,
\begin{equation}
\frac{ -i\langle T a^{t}_{ {\bf{k}} }({\bf{q}})
a^{\dagger}_{ {\bf{k}}^{'} }({\bf{q}}^{'}) \rangle }{\langle T 1 \rangle}
 = \sum_{n}\mbox{ }exp(\omega_{n}t)\mbox{ }
\frac{ -i\langle T a^{n}_{ {\bf{k}} }({\bf{q}})
a^{\dagger}_{ {\bf{k}}^{'} }({\bf{q}}^{'}) \rangle }{\langle T 1 \rangle}
\end{equation}
\begin{equation}
\frac{ -i\langle T a^{\dagger t}_{ {\bf{k}} }(-{\bf{q}})
a^{\dagger}_{ {\bf{k}}^{'} }({\bf{q}}^{'}) \rangle }{\langle T 1 \rangle}
 = \sum_{n}\mbox{ }exp(\omega_{n}t)\mbox{ }
\frac{ -i\langle T a^{\dagger n}_{ {\bf{k}} }(-{\bf{q}})
a^{\dagger}_{ {\bf{k}}^{'} }({\bf{q}}^{'}) \rangle }{\langle T 1 \rangle}
\end{equation}
and, $ \omega_{n} = (2\mbox{ }\pi \mbox{ }n)/\beta $.  Thus,
\[
(i\omega_{n} - \omega_{ {\bf{k}} }({\bf{q}}))
\frac{ -i\langle T a^{n}_{ {\bf{k}} }({\bf{q}})
a^{\dagger}_{ {\bf{k}}^{'} }({\bf{q}}^{'}) \rangle }{\langle T 1 \rangle}
 = \frac{ \delta_{ {\bf{k}}, {\bf{k}}^{'} }
\delta_{ {\bf{q}}, {\bf{q}}^{'} } }{-i \mbox{ }\beta}
\]
\begin{equation}
+ (\frac{ v_{ {\bf{q}} } }{V})
\Lambda_{ {\bf{k}} }(-{\bf{q}})
\sum_{ {\bf{k}}^{''} }[\Lambda_{ {\bf{k}}^{''} }(-{\bf{q}})
\frac{ -i\langle T a^{n}_{ {\bf{k}}^{''} }({\bf{q}})
a^{\dagger}_{ {\bf{k}}^{'} }({\bf{q}}^{'}) \rangle }{\langle T 1 \rangle}
+ \Lambda_{ {\bf{k}}^{''} }({\bf{q}})
\frac{ -i\langle T a^{\dagger n}_{ {\bf{k}}^{''} }(-{\bf{q}})
a^{\dagger}_{ {\bf{k}}^{'} }({\bf{q}}^{'}) \rangle }{\langle T 1 \rangle}]
\end{equation}
\[
(i\omega_{n} + \omega_{ {\bf{k}} }(-{\bf{q}}))
\frac{ -i\langle T a^{\dagger n}_{ {\bf{k}} }(-{\bf{q}})
a^{\dagger}_{ {\bf{k}}^{'} }({\bf{q}}^{'}) \rangle }{\langle T 1 \rangle}
\]
\begin{equation}
= -(\frac{ v_{ {\bf{q}} } }{V})
\Lambda_{ {\bf{k}} }({\bf{q}})
\sum_{ {\bf{k}}^{''} }[\Lambda_{ {\bf{k}}^{''} }({\bf{q}})
\frac{ -i\langle T a^{\dagger n}_{ {\bf{k}}^{''} }(-{\bf{q}})
a^{\dagger}_{ {\bf{k}}^{'} }({\bf{q}}^{'}) \rangle }{\langle T 1 \rangle}
+ \Lambda_{ {\bf{k}}^{''} }(-{\bf{q}})
\frac{ -i\langle T a^{n}_{ {\bf{k}}^{''} }({\bf{q}})
a^{\dagger}_{ {\bf{k}}^{'} }({\bf{q}}^{'}) \rangle }{\langle T 1 \rangle} ]
\end{equation}
Define,
\begin{equation}
\sum_{ {\bf{k}} } \Lambda_{ {\bf{k}} } (-{\bf{q}})
\frac{ -i\langle T a^{n}_{ {\bf{k}} }({\bf{q}})
a^{\dagger}_{ {\bf{k}}^{'} }({\bf{q}}^{'}) \rangle }{\langle T 1 \rangle}
 = G_{1}( {\bf{q}}, {\bf{k}}^{'}, {\bf{q}}^{'}; n)
\end{equation}
\begin{equation}
\sum_{ {\bf{k}} } \Lambda_{ {\bf{k}} } ({\bf{q}})
\frac{ -i\langle T a^{\dagger n}_{ {\bf{k}} }(-{\bf{q}})
a^{\dagger}_{ {\bf{k}}^{'} }({\bf{q}}^{'}) \rangle }{\langle T 1 \rangle}
 = G_{2}( {\bf{q}}, {\bf{k}}^{'}, {\bf{q}}^{'}; n)
\end{equation}
Multiplying the above equations with $ \Lambda_{ {\bf{k}} }(-{\bf{q}}) $ and
summing over $ {\bf{k}} $ one arrives at simple formulas for
 $ G_{1} $ and $ G_{2} $.
\[
G_{1}( {\bf{q}}, {\bf{k}}^{'}, {\bf{q}}^{'}; n) =
\Lambda_{ {\bf{k}}^{'} }(-{\bf{q}})
\frac{ \delta_{ {\bf{q}}, {\bf{q}}^{'} } }
{ -i \mbox{ }\beta( i\omega_{n} - \omega_{ {\bf{k}}^{'} }({\bf{q}}) ) }
\]
\begin{equation}
+ f_{n}({\bf{q}})[G_{1}( {\bf{q}}, {\bf{k}}^{'}, {\bf{q}}^{'}; n)
+ G_{2}( {\bf{q}}, {\bf{k}}^{'}, {\bf{q}}^{'}; n) ]
\end{equation}
and,
\begin{equation}
G_{2}( {\bf{q}}, {\bf{k}}^{'}, {\bf{q}}^{'}; n) =
f^{*}_{n}(-{\bf{q}})[ G_{1}( {\bf{q}}, {\bf{k}}^{'}, {\bf{q}}^{'}; n)
+ G_{2}( {\bf{q}}, {\bf{k}}^{'}, {\bf{q}}^{'}; n) ]
\end{equation}
\[
G_{ 2 }({\bf{q}}, {\bf{k}}^{'}, {\bf{q}}^{'}; n) =
  \frac{ f^{*}_{n}(-{\bf{q}}) }{(1- f^{*}_{n}(-{\bf{q}}))}
G_{1} ({\bf{q}}, {\bf{k}}^{'}, {\bf{q}}^{'}; n)
\]
\[
 G_{1}({\bf{q}}, {\bf{k}}^{'}, {\bf{q}}^{'}; n) +
 G_{2}({\bf{q}}, {\bf{k}}^{'}, {\bf{q}}^{'}; n)
 = G_{1}({\bf{q}}, {\bf{k}}^{'}, {\bf{q}}^{'}; n) /( 1 - f^{*}_{n}(-{\bf{q}}) )
\]
\begin{equation}
G_{1}({\bf{q}}, {\bf{k}}^{'}, {\bf{q}}^{'}; n)
 = (\frac{1}{-i \mbox{ }\beta})
\frac{ (1-f_{n}^{*}(-{\bf{q}}) ) \Lambda_{ {\bf{k}}^{'} }(-{\bf{q}})
\delta_{ {\bf{q}}, {\bf{q}}^{'} } }
{(1- f_{n}^{*}(-{\bf{q}}) - f_{n}({\bf{q}}) )
 ( i\omega_{n} - \omega_{ {\bf{k}}^{'} }({\bf{q}}) ) }
\end{equation}
\begin{equation}
G_{2}({\bf{q}}, {\bf{k}}^{'}, {\bf{q}}^{'}; n)
 = (\frac{1}{-i \mbox{ }\beta})
\frac{ f_{n}^{*}(-{\bf{q}}) \Lambda_{ {\bf{k}}^{'} }(-{\bf{q}})
\delta_{ {\bf{q}}, {\bf{q}}^{'} } }
{(1- f_{n}^{*}(-{\bf{q}}) - f_{n}({\bf{q}}) )
 ( i\omega_{n} - \omega_{ {\bf{k}}^{'} }({\bf{q}}) ) }
\end{equation}
\[
G_{1}({\bf{q}}, {\bf{k}}^{'}, {\bf{q}}^{'}; n)
 + G_{2}({\bf{q}}, {\bf{k}}^{'}, {\bf{q}}^{'}; n)
 = (\frac{1}{-i \mbox{ }\beta})\frac{ \Lambda_{ {\bf{k}}^{'} }(-{\bf{q}})
\delta_{ {\bf{q}}, {\bf{q}}^{'} } }
{(1- f_{n}^{*}(-{\bf{q}}) - f_{n}({\bf{q}}) )
 ( i\omega_{n} - \omega_{ {\bf{k}}^{'} }({\bf{q}}) ) }
\]

\[
\frac{ -i\langle T a^{n}_{ {\bf{k}} }({\bf{q}})
a^{\dagger}_{ {\bf{k}}^{'} }({\bf{q}}^{'}) \rangle }{ \langle T 1 \rangle }
 = \frac{ \delta_{ {\bf{k}}, {\bf{k}}^{'} } \delta_{ {\bf{q}}, {\bf{q}}^{'} } }
{-i\mbox{ }\beta( i\omega_{n} - \omega_{ {\bf{k}} }({\bf{q}}) )}
\]
\begin{equation}
+ (\frac{1}{-i \mbox{ }\beta})
( \frac{ v_{ {\bf{q}} } }{V} )\frac{ \Lambda_{ {\bf{k}} }(-{\bf{q}}) }
{(i\omega_{ n } - \omega_{ {\bf{k}} }({\bf{q}}) ) }
\frac{ \Lambda_{ {\bf{k}}^{'} }(-{\bf{q}})
\delta_{ {\bf{q}}, {\bf{q}}^{'} } }
{(1- f_{n}^{*}(-{\bf{q}}) - f_{n}({\bf{q}}) )
 ( i\omega_{n} - \omega_{ {\bf{k}}^{'} }({\bf{q}}) ) }
\end{equation}
The zero temperature correlation function of significance here is,
\begin{equation}
-i\langle a^{\dagger}_{ {\bf{k}}^{'} }({\bf{q}}^{'})
a_{ {\bf{k}} }({\bf{q}}) \rangle
\end{equation}
This may be obtained from the above formulas as,
\begin{equation}
-i\langle a^{\dagger}_{ {\bf{k}}^{'} }({\bf{q}}^{'})
a_{ {\bf{k}} }({\bf{q}}) \rangle
 = -(\frac{v_{ {\bf{q}} } }{V})
\Lambda_{ {\bf{k}} }(-{\bf{q}})\Lambda_{ {\bf{k}}^{'} }(-{\bf{q}})
\delta_{ {\bf{q}}, {\bf{q}}^{'} }
\int_{C} \frac{d\omega}{2\pi \mbox{ }i}
\frac{1}{( i\omega - \omega_{ {\bf{k}} }({\bf{q}}) )
( i\omega - \omega_{ {\bf{k}}^{'} }({\bf{q}}) )
 ( 1 - f_{n}^{*}(-{\bf{q}}) - f_{n}({\bf{q}}) ) }
\end{equation}
 where $ C $ is the positively oriented contour
 that encloses the upper half-plane( upper half-plane, because we need
 $ \langle a^{\dagger}_{ {\bf{k}}^{'} }({\bf{q}}^{'})a_{ {\bf{k}} }({\bf{q}}) \rangle $ and not
 $ \langle a_{ {\bf{k}} }({\bf{q}})a^{\dagger}_{ {\bf{k}}^{'} }({\bf{q}}^{'})\rangle $ )
 Thus the problem now reduces to computing all the zeros
 of $ ( 1 - f_{n}^{*}(-{\bf{q}}) - f_{n}({\bf{q}}) )  $
 that have positive imaginary parts. It may be shown quite easily that,
\begin{equation}
\epsilon_{RPA}({\bf{q}}, i\omega_{n}) = 
1 - f_{n}^{*}(-{\bf{q}}) - f_{n}({\bf{q}})
\end{equation}
In 1D, the dielectric function is evaluated as follows,
\begin{equation}
1 - f^{*}_{n}(-q) - f_{n}(q) =
1 + v_{q}(\frac{1}{2\pi})
(\frac{m}{q})ln[ \frac{ (k_{f}+q/2)^{2} + (\frac{m \mbox{ }\omega}{q})^{2} }
{ (k_{f} - q/2)^{2} + (\frac{m \mbox{ }\omega}{q})^{2} } ]
 = 0
\end{equation}
This leads to the root,
\begin{equation}
\omega = i\mbox{ }(\frac{|q|}{m})
\sqrt{ \frac{ (k_{f}+q/2)^{2} - (k_{f} - q/2)^{2}
exp(-(\frac{ 2\mbox{ }\pi\mbox{ }q }{m})(\frac{1}{v_{q}})) }
{ 1- exp(-(\frac{ 2\mbox{ }\pi\mbox{ }q }{m})(\frac{1}{v_{q}})) } }
\label{ROOT}
\end{equation}
Therefore the final result may be written as,
\begin{equation}
\langle a^{\dagger}_{ {\bf{k}}^{'} }({\bf{q}}^{'})a_{ {\bf{k}} }({\bf{q}})
 \rangle
 = (\frac{1}{V})\frac
{ \Lambda_{k}(-q)\Lambda_{k^{'}}(-q) \delta_{ q, q^{'} } }
{ (\omega_{R}(q)+ \omega_{k}(q))(\omega_{R}(q)+ \omega_{k^{'}}(q))
(\frac{m}{q^{2}})(\frac{1}{2 \pi k_{f}})2(\frac{m}{q})^{2}\omega_{R}(q)
(cosh(\lambda(q))-1) }
\end{equation}
here,
\begin{equation}
\lambda(q) = (\frac{2 \pi q}{m})(\frac{1}{v_{q}})
\end{equation}
\begin{equation}
\omega_{R}(q) = (\frac{ |q| }{m})
\sqrt{ \frac{ (k_{f} + q/2)^{2}  - (k_{f} - q/2)^{2}exp(-\lambda(q)) }
{ 1 - exp(-\lambda(q)) } }
\end{equation}
In other words,
\[
\langle c^{\dagger}_{ k }c_{ k } \rangle
 = n_{F}(k) +
(2\pi k_{f})
\int_{-\infty}^{+\infty} \mbox{ }\frac{ dq_{1} }{2\pi}\mbox{ }
\frac{ \Lambda_{ k - q_{1}/2 }(-q_{1}) }
{ 2\omega_{R}(q_{1})(\omega_{R}(q_{1}) + \omega_{k - q_{1}/2}(q_{1}))^{2}
(\frac{ m^{3} }{q_{1}^{4}})( cosh(\lambda(q_{1})) - 1 ) }
\]
\begin{equation}
- (2\pi k_{f})
\int_{-\infty}^{+\infty} \mbox{ }\frac{ dq_{1} }{2\pi}\mbox{ }
\frac{ \Lambda_{ k + q_{1}/2 }(-q_{1}) }
{ 2\omega_{R}(q_{1})(\omega_{R}(q_{1}) + \omega_{k + q_{1}/2}(q_{1}))^{2}
(\frac{ m^{3} }{q_{1}^{4}})( cosh(\lambda(q_{1})) - 1 ) }
\end{equation}
 Note that the above formula possesses a non-analytic dependence
 in the coupling strength
 $ ( cosh(\frac{2 \pi q}{m})(\frac{1}{v_{q}}) - 1 ) $ ,
 an unmistakeable signature of a non-digrammatic result.
 Next, we would like to provide formulas for the momentum distribution when
 we use the correct interacting expectation values in the 
 fermi-bilinear sea-boson correspondence. The results obtained
 from these formulas are likely
 to be very different from the weakly nonideal case, which in any case
 is not very interesting. The answers are given below,
\begin{equation}
{\bar{n}}_{ {\bf{k}} } = 
\frac{ n^{\beta}({\bf{k}}) }{S_{1}({\bf{k}}) }
+ \frac{ S_{2}({\bf{k}}) }{S_{1}({\bf{k}}) }
\end{equation}
here,
\begin{equation}
S_{1}({\bf{k}}) = 1 + \sum_{ {\bf{q}},i }
\frac{ {\bar{n}}_{ {\bf{k}} - {\bf{q}} } }
{({\tilde{\omega}}_{i}(-{\bf{q}}) + \frac{ {\bf{k.q}} }{m} - \epsilon_{ {\bf{q}} })^{2}}
g^{2}_{i}(-{\bf{q}})
 + \sum_{ {\bf{q}},i }
\frac{ 1-{\bar{n}}_{ {\bf{k}} + {\bf{q}} } }
{({\tilde{\omega}}_{i}(-{\bf{q}}) + \frac{ {\bf{k.q}} }{m} + \epsilon_{ {\bf{q}} })^{2}}
g^{2}_{i}(-{\bf{q}})
\end{equation}
\begin{equation}
S_{2}({\bf{k}}) = \sum_{ {\bf{q}},i }
\frac{ {\bar{n}}_{ {\bf{k}} - {\bf{q}} } }
{({\tilde{\omega}}_{i}(-{\bf{q}}) + \frac{ {\bf{k.q}} }{m} - \epsilon_{ {\bf{q}} })^{2}}
g^{2}_{i}(-{\bf{q}})
\end{equation}
also the form of the "RPA" dielectric function and its zeros
 $ {\tilde{\omega}}_{i}({\bf{q}}) $ are now different. The "RPA" dielectric
 function is given by,
\begin{equation}
\epsilon_{RPA}({\bf{q}},\omega) = 1 + \frac{ v_{{\bf{q}}} }{V}
\sum_{ {\bf{k}} }\frac{ {\bar{n}}_{ {\bf{k}} + {\bf{q}}/2 }
 - {\bar{n}}_{ {\bf{k}} - {\bf{q}}/2 } }
{\omega - \frac{ {\bf{k.q}} }{m}}
\end{equation}
\begin{equation}
g_{i}({\bf{q}})
 = [\sum_{ {\bf{k}} }
\frac{ {\bar{n}}_{ {\bf{k}} - {\bf{q}}/2 }-{\bar{n}}_{ {\bf{k}} + {\bf{q}}/2 } }
{({\tilde{\omega}}_{i}({\bf{q}}) - \frac{ {\bf{k.q}} }{m})^{2}}]
^{-\frac{1}{2}}
\end{equation}
The commutators are given as before, except for three changes. In the new 
approach,
\begin{equation} 
 \Lambda_{ {\bf{k}} }({\bf{q}}) = \sqrt{ {\bar{n}}_{ {\bf{k}}+{\bf{q}}/2 }
(1-{\bar{n}}_{ {\bf{k}}-{\bf{q}}/2 }) }
\end{equation}
 Next, the zeros are slightly different. The collective mode has to be
 computed self-consistently, whereas the particle-hole mode may be written
 down as described earlier,
\begin{equation}
 {\tilde{\omega}}_{i}( {\bf{q}} ) = \omega_{ {\bf{k}}_{i} }({\bf{q}})
 + (\frac{ v_{ {\bf{q}} } }{V})
\frac{ \Lambda^{2}_{ {\bf{k}}_{i} }(-{\bf{q}}) }
{ 1 + \frac{ v_{ {\bf{q}} } }{V}
\sum_{ {\bf{k}} }\frac{ \Lambda^{2}_{ {\bf{k}} }({\bf{q}}) }
{ \omega_{ {\bf{k}}_{i} }({\bf{q}}) + \omega_{ {\bf{k}} }(-{\bf{q}}) }
- \frac{ v_{ {\bf{q}} } }{V}
\sum_{ {\bf{k}} \neq {\bf{k}}_{i}}\frac{ \Lambda^{2}_{ {\bf{k}} }(-{\bf{q}}) }
{ \omega_{ {\bf{k}}_{i} }({\bf{q}}) - \omega_{ {\bf{k}} }({\bf{q}}) } }
\end{equation}
 The last change is in the form of $ U_{0}({\bf{q}}) $, here we have
 to make sure we use the finite temperature noninteracting values.
 The other issue that is also of interest is whether the momentum distribution
 evaluated using the fermi-bilinear/sea-boson correspondence is the same
 as that suggested by the full propagator. We have found that the answer
 to this is difficult and probably in the negative. This does not
 mean that the whole program is wrong. Some comfort and confidence
 in these manipulations may still be retained by demonstrating that
 the expression for the number operator is consistent with
 the RPA-form of the fermi creation operator. Again here, we have
 to be content with a weak form of this requirement. We take
 the point of view that it is sufficient to show that the commutator between
 the total momentum of the electrons and the field operator comes out
 the same in both the original fermi language and in the sea-boson
 language. The total momentum of the electrons has the form,
\begin{equation}
{\bf{P}} = \sum_{ {\bf{k}} }{\bf{k}}
\mbox{            }c^{\dagger}_{ {\bf{k}} }c_{ {\bf{k}} }
\end{equation}
In the sea-boson language, it takes the form,
\begin{equation}
{\bf{P}} = \sum_{ {\bf{k}}, {\bf{q}} }{\bf{q}}
\mbox{            }a^{\dagger}_{ {\bf{k}} }({\bf{q}})
a_{ {\bf{k}} }({\bf{q}})
\end{equation}
Therefore in the original fermi language we have,
\begin{equation}
[{\bf{P}}, \psi^{\dagger}({\bf{x}})] = i\mbox{        }\nabla
\mbox{        }\psi^{\dagger}({\bf{x}})
\end{equation}
In the sea-boson language we have,
\[
[{\bf{P}}, \psi^{\dagger}({\bf{x}})] =
 \sum_{ {\bf{k}}, {\bf{q}} }{\bf{q}}
[a^{\dagger}_{ {\bf{k}} }({\bf{q}})a_{ {\bf{k}} }({\bf{q}}),
\psi^{\dagger}({\bf{x}})]
\]
\[
 = \sum_{ {\bf{k}}, {\bf{q}} }{\bf{q}}a^{\dagger}_{ {\bf{k}} }({\bf{q}})
(-g_{ {\bf{k}}, {\bf{q}} }({\bf{x}}))\psi^{\dagger}({\bf{x}})
+ \sum_{ {\bf{k}}, {\bf{q}} }{\bf{q}}
(-g^{*}_{ {\bf{k}}, {\bf{q}} }({\bf{x}})) \psi^{\dagger}({\bf{x}})
a_{ {\bf{k}} }({\bf{q}})
 = i\mbox{   }\nabla\mbox{   }\psi^{\dagger}({\bf{x}})
\]
 as it should be.
 All this points to the fact that the answers for the momentum
 distribution and propagators should not be taken too
 literally, rather one must be content with the qualitative predictions
 that are most likely accurate, that also seem to contradict
 conventional wisdom.

\begin{center} APPENDIX E \end{center}

 In the late 70's and early 80's, attempts were made to write down field
 theories that describe scalar mesons in terms of observables like
 currents and densities rather than the creation and annhilation operators.
 The motivation for doing this stems from the fact a theory cast directly
 in terms of observables was more physically intuitive than the 
 more traditional
 approach based on raising and lowering operators on the Fock space.
 This attempt however, raised a number of technical questions, among them
 was how to make sense of the various identities connecting say the kinetic
 energy density to the currents and particle densities and so on.
 Elaborate mathematical machinery was erected by the authors who started
 this line of research\cite{Sharp} to address these issues. However, it seems
 gaps still remain especially with regard to the crucial question of how
 one goes about writing down a formula for the annhilation operator (fermi or
 bose) alone in terms of bilinears like currents and densities. The
 bilinears in question namely currents and densities satisfy a closed algebra
 known as the current algebra \cite{Sharp}. This algebra is insensitive
 to the nature of the statistics of the underlying fields. On the other hand,
 if one desires information about single-particle properties, it is necessary
 to relate the annhilation operator (whose commutation rules determine the
 statistics) to bilinears like currents and densities. That such a
 correspondence is possible was demonstrated by Goldin, Menikoff and Sharp
 \cite{Sharp}. However, they have not explicitly written down such a formula
 nor have they clarified some important issues such as whether this formula
 changes when one consider interacting fields rather than free fields.
 The general belief\cite{REFEREE} is that these formulas are different
 for interacting fields. It is shown here that this is in fact not the
 case, interactions in the system merely cause a change
 in the hamiltonian but do not affect how the annhilation operator is
 related to local currents and densities.
  The attempts made here are partly based on the work of Goldin et.al.
  \cite{Sharp}  Ligouri and
  Mintchev on generalised statistics\cite{Lig} and the series by
  Reed and Simon\cite{Reed}.
  As has been demonstrated earlier, for the bose case we had to choose 
 $ \Phi = 0 $. We argued that this choice was unique. In the fermi case
 the choice was different but was also unique due to the necessity of 
 recovering the free theory. In this section, we write down a mathematically
 rigorous statement of this uniqueness criterion.
 This exercise also settles
 the issue regarding the delicate question of multiplying two operator-valued
 distributions at the same point and other related issues, like the meaning
 of the square-root of the density distribution.
 For this we prove this lemma.
\newline
{\bf{Lemma}}
 Let $ {\mathcal{F}} $ be smooth a function from a bounded
 subset of the real line on to the set of reals. Also let
 $ f $ and $ g $ be smooth functions from some bounded subset of
 $ {\mathcal{R}}^{d} $ to reals.
 Let us further assume that
 the range of these functions are such that it is always possible to
 find compositions such as $ {\mathcal{F}}\mbox{ }o\mbox{ }f $
 and they will also be smooth functions with sufficiently big domains.
 They possess fourier transforms since they are well-behaved. 
 If,
\begin{equation}
{\mathcal{F}}(\mbox{  }f({\bf{x}})\mbox{  }) = g({\bf{x}})
\end{equation}
and,
\begin{equation}
f({\bf{x}}) = \sum_{ {\bf{q}} }{\tilde{f}}_{ {\bf{q}} }
\mbox{  }e^{ i\mbox{ }{\bf{q.x}} }
\end{equation}
\begin{equation}
g({\bf{x}}) = \sum_{ {\bf{k}} }{\tilde{g}}_{ {\bf{k}} }
\mbox{  }e^{ i\mbox{ }{\bf{k.x}} }
\end{equation}
then the following also holds,
\begin{equation}
[ {\mathcal{F}}(\mbox{ }\sum_{ {\bf{q}} }
{\tilde{f}}_{ {\bf{q}} }T_{ -{\bf{q}} }({\bf{k}}) \mbox{ })]
\mbox{ }\delta_{ {\bf{k}}, {\bf{0}} } = {\tilde{g}}_{ {\bf{k}} }
\end{equation}
where $ T_{ {\bf{q}} }({\bf{k}}) = exp({\bf{q}}.\nabla_{ {\bf{k}} }) $.
Here the operator $ T_{ {\bf{q}} }({\bf{k}}) $ acts on the $ {\bf{k}} $
 in the Kronecker delta on the extreme right, and every time it translates
 the $ {\bf{k}} $ by an amount $ {\bf{q}} $.
\newline
{\bf{Proof}}
\newline
Proof is by brute force expansion.
We know,
\begin{equation}
{\mathcal{F}}(y) = \sum_{n=0}^{\infty} \frac{ {\mathcal{F}}^{(n)}(0) }{n!}
\mbox{ }y^{n}
\end{equation}
therefore,
\[
{\mathcal{F}}( \mbox{  }f({\bf{x}}) \mbox{  })
 = {\mathcal{F}}(0) + \sum_{n=1}^{\infty} \frac{ {\mathcal{F}}^{(n)}(0) }{n!}
\mbox{ }\sum_{ \{ {\bf{q}}_{i} \} }{\tilde{f}}_{ {\bf{q}}_{1} }
{\tilde{f}}_{ {\bf{q}}_{2} }...{\tilde{f}}_{ {\bf{q}}_{n} }
exp(i(\sum_{i=1}^{n}{\bf{q}}_{i}).{\bf{x}})
\]
\begin{equation}
= \sum_{ {\bf{k}} }e^{i\mbox{ }{\bf{k.x}} }{\tilde{g}}_{ {\bf{k}} }
\end{equation}
This means (take inverse fourier transform),
\[
 {\mathcal{F}}(0)\delta_{ {\bf{k}}, 0 }
 + \sum_{n=1}^{\infty} \frac{ {\mathcal{F}}^{(n)}(0) }{n!}
\mbox{ }\sum_{ \{ {\bf{q}}_{i} \} }{\tilde{f}}_{ {\bf{q}}_{1} }
{\tilde{f}}_{ {\bf{q}}_{2} }...{\tilde{f}}_{ {\bf{q}}_{n} }
\delta_{ ( {\bf{k}} - \sum_{i=1}^{n}{\bf{q}}_{i} ), {\bf{0}} }
\]
\begin{equation}
= {\tilde{g}}_{ {\bf{k}} }
\end{equation}
This may also be cleverly rewritten as
\[
 {\mathcal{F}}(0)\delta_{ {\bf{k}}, 0 }
 + \sum_{n=1}^{\infty} \frac{ {\mathcal{F}}^{(n)}(0) }{n!}
\mbox{ }\sum_{ \{ {\bf{q}}_{i} \} }{\tilde{f}}_{ {\bf{q}}_{1} }
{\tilde{f}}_{ {\bf{q}}_{2} }...{\tilde{f}}_{ {\bf{q}}_{n} }
T_{ -{\bf{q}}_{1} }({\bf{k}})T_{ -{\bf{q}}_{2} }({\bf{k}})
...T_{ -{\bf{q}}_{n} }({\bf{k}})
\delta_{  {\bf{k}}, {\bf{0}} }
\]
\begin{equation}
= {\tilde{g}}_{ {\bf{k}} }
\end{equation}
and therefore,
\begin{equation}
{\tilde{g}}_{ {\bf{k}} } = [{\mathcal{F}}
(\mbox{  }\sum_{ {\bf{q}} }{\tilde{f}}_{ {\bf{q}} }
T_{ -{\bf{q}} }({\bf{k}}) \mbox{  } )]\mbox{  }\delta_{ {\bf{k}}, {\bf{0}} }
\end{equation}
and the {\bf{Proof}} is now complete.

 Now we would like to capture the notion of the fermi density operator.
 Physicists define it to be $ \rho(x) = \psi^{*}(x) \psi(x) $. Multiplication
 of two fermi fields at the same point is a delicate issue and we would
 like to make more sense out of it. For this we have to set our single
 particle Hilbert Space:
\[
 {\mathcal{H}} = L_{p}({\mathcal{R}}^{3}) {\bigotimes} {\mathcal{W}}
\]
 Here, $ L_{p}({\mathcal{R}}^{3}) $ is the space of all periodic functions
 with period $ L $ in each space direction. $ {\mathcal{W}} $
 is the spin space spanned by two vectors.
 An orthonormal basis for $ {\mathcal{W}} $
\[
\{ \xi_{\uparrow}, \xi_{\downarrow} \}
\]
 A typical element of $ {\mathcal{H}} $ is given by
 $ f({\bf{x}}) \bigotimes \xi_{\downarrow} $. A basis for $ {\mathcal{H}} $
 is given by
\[
{\mathcal{B}} =
 \{ \sqrt{ \frac{1}{L^{3}} } exp(i{\bf{q_{n}.x}}) \bigotimes
 \xi_{s}
  ;
 {\bf{n}} = (n_{1}, n_{2}, n_{3}) \in {\mathcal{Z}}^{3},
  s \in \{ \uparrow, \downarrow \} ;
\]
 We move on to the definition of the fermi-density distribution.
 The Hilbert Space $ {\mathcal{H}}^{\bigotimes n} $ is the space of
 all n-particle wavefunctions with no symmetry restrictions.
 From this we may construct orthogonal subspaces
\[
{\mathcal{H}}_{+}^{\bigotimes n} = P_{+} {\mathcal{H}}^{\bigotimes n}
\]
\[
{\mathcal{H}}_{-}^{\bigotimes n} = P_{-} {\mathcal{H}}^{\bigotimes n}
\]
 Tensors from $ {\mathcal{H}}_{+}^{\bigotimes n} $ are orthogonal
 to tensors from $ {\mathcal{H}}_{-}^{\bigotimes n} $.
 The only exceptions are when $ n = 0 $ or $ n = 1 $.
 The space $ {\mathcal{H}}_{+}^{\bigotimes n} $ is the space of
 bosonic-wavefunctions and the space $ {\mathcal{H}}_{-}^{\bigotimes n} $
 is the space of fermionic wavefunctions. The definition of the fermi
 density distribution proceeds as follows. Let $ v $ be written as
\[
 v = \sum_{\sigma \in \{ \uparrow, \downarrow \} }a(\sigma)\xi_{\sigma}
\]
 The Fermi density distibution is an operator on the Fock Space,
 given a vector $ f \bigotimes v \in \mathcal{H} $ in the single particle
 Hilbert Space, and a tensor $ \varphi $ in the n-particle subspace of
 of $ \mathcal{F}(\mathcal{H}) $, there exists a corresponding operator
 $ \rho(f \bigotimes v) $ that acts as follows:
\[
[\rho(f \bigotimes v) \varphi ]_{n}
({\bf{x_{1}}}\sigma_{1}, {\bf{x_{2}}}\sigma_{2},
 .... , {\bf{x_{n}}}\sigma_{n})
  = 0
\]
  if $ \varphi \in  {\mathcal{H}}_{+}^{\bigotimes n} $ and
\[
[\rho(f \bigotimes v) \varphi]_{n}
 ({\bf{x_{1}}}\sigma_{1}, {\bf{x_{2}}}\sigma_{2},
 .... , {\bf{x_{n}}}\sigma_{n})
 = \sum_{i=1}^{n} f({\bf{x_{i}}}) a(\sigma_{i})
 \varphi_{n}({\bf{x_{1}}}\sigma_{1}, {\bf{x_{2}}}\sigma_{2},
 .... , {\bf{x_{n}}}\sigma_{n})
\]
when $ \varphi \in  {\mathcal{H}}_{-}^{\bigotimes n} $.
The physical meaning of this abstract operator will become clear
 soon.
Let us now define the current density in an analogous fashion,
To Physicists, it is,
\begin{equation}
{\bf{J}}({\bf{x}}) = (\frac{1}{2i})[ \psi^{\dagger}(\nabla\psi)
 - (\nabla\psi)^{\dagger}\psi ]
\end{equation}
 To Mathematicians it is an operator similar to the density
\cite{Lig}. Given a typical element
 $ f \bigotimes v $ associated with the underlying single-particle
 Hilbert space, there is an operator denoted
 by $ J_{s}(f \bigotimes v) $, ( $ s = 1,2,3 $ )
 that acts on a typical tensor from the
 n-particle subspace of the full Fock space as follows,
\begin{equation}
 [J_{s}(f \bigotimes v)\varphi]_{n}
({\bf{x}}_{1}\sigma_{1}, {\bf{x}}_{2}\sigma_{2}, ... , {\bf{x}}_{n}\sigma_{n})
 = 0
\end{equation}
if $ \varphi \in {\mathcal{H}}^{\bigotimes n}_{+} $,and
and,
\[
 [J_{s}(f \bigotimes v)\varphi]_{n}
({\bf{x}}_{1}\sigma_{1}, {\bf{x}}_{2}\sigma_{2}, ... , {\bf{x}}_{n}\sigma_{n})
\]
\begin{equation}
 = -i\sum_{k = 1}^{n}
\{ f({\bf{x}}_{k})a(\sigma_{k})\nabla^{k}_{s} + \frac{1}{2}
[\nabla^{k}_{s}f({\bf{x}}_{k})]a(\sigma_{k}) \} \varphi_{n}
({\bf{x}}_{1}\sigma_{1}, {\bf{x}}_{2}\sigma_{2}, ... , {\bf{x}}_{n}\sigma_{n})
\end{equation}
if $ \varphi \in {\mathcal{H}}^{\bigotimes n}_{-} $.
For the bosonic current it is the other way around.
Having done all this, we would now like to write the DPVA more
 rigorously.
Now for some notation. As before, let
 $ g = \mbox{ }exp(i{\mbox{ }}{\bf{k}}_{m}.{\bf{x}})\bigotimes \xi_{r} $
(the square root of the volume is not needed as we want all operators
in momentum space to be dimensionless).
Then as before,
\begin{equation}
\psi({\bf{k}}_{m}r) = c(g)
\end{equation}
\begin{equation}
\rho({\bf{k}}_{m}r) = \rho(g)
\end{equation}
\begin{equation}
\delta \rho({\bf{k}}_{m}r) = \rho({\bf{k}}_{m}r) - N^{0}_{r}
\delta_{ {\bf{k}}_{m}, {\bf{0}} }
\end{equation}
\begin{equation}
j_{s}({\bf{k}}_{m}r) = {\bf{J}}_{s}(g)
\end{equation}
\begin{equation}
\delta j_{s}({\bf{k}}_{m}r) = j_{s}({\bf{k}}_{m}r)
\end{equation}
Having done this, we would like to write down another formula for the canonical
conjugate. 
\begin{equation}
\nabla\Pi({\bf{x}}\sigma) = (-1/\rho({\bf{x}}\sigma)){\bf{J}}({\bf{x}}\sigma)
+ \nabla\Phi([\rho];{\bf{x}}\sigma) - [-i\mbox{ }\Phi,\nabla\Pi]
\end{equation}
 Then we have(bear in mind here that we have distinguished between the c-number
 $ N^{0}_{r} $ and the operator $ \rho({\bf{0}}r) $ whose expectation value
 is $ N^{0}_{r} $).
\begin{equation}
(i\mbox{ }{\bf{q}}_{m})\mbox{  }X_{ {\bf{q}}_{m}r }
 = -(\frac{1}{ N^{0}_{r} })\frac{1}{1 + \frac{1}{N^{0}_{r}}
\sum_{ {\bf{k}}_{n} }
\delta\rho({\bf{k}}_{n}r)T_{ {\bf{k}}_{n} }({\bf{q}}_{m} )}
[ \mbox{  }
\sum_{ {\bf{p}}_{n} }\delta {\bf{j}}({\bf{p}}_{n}r)
T_{ {\bf{p}}_{n} }({\bf{q}}_{m} )
\mbox{   } ] \mbox{   } \delta_{ {\bf{q}}_{m}, {\bf{0}} }
+ \mbox{    }{\bf{F}}([\rho];{\bf{q}}_{m}r)
\end{equation}
where,
\begin{equation}
\sum_{ {\bf{q}}_{m} }exp(i{\mbox{  }}{\bf{q}}_{m}.{\bf{x}})
{\bf{F}}([\rho];{\bf{q}}_{m}r)
 = \nabla\Phi - [-i\mbox{ }\Phi,\nabla\Pi]
\end{equation}
 As regards the object $ X_{ {\bf{0}}r } $ that is conjugate to the
 total number is concerned, we must retain it as it is, since,
 it will ensure that the total number when commuting with the field 
 operator is the field operator itself rather than the incorrect answer
 zero.
 For
 $ {\bf{q}}_{m} \neq {\bf{0}} $
\[
X_{ {\bf{q}}_{m}r }
 = (\frac{1}{q_{m}^{2}})(\frac{ i }{N^{0}_{r}})
\frac{1}{1 + \frac{1}{N^{0}_{r}}
\sum_{ {\bf{k}}_{n} }
\delta\rho({\bf{k}}_{n}r)T_{ {\bf{k}}_{n} }({\bf{q}}_{m} )}
[ \mbox{  }
\sum_{ {\bf{p}}_{n} }
{\bf{q}}_{m}.\delta {\bf{j}}({\bf{p}}_{n}r)
T_{ {\bf{p}}_{n} }({\bf{q}}_{m} )
\mbox{   } ] \mbox{   } \delta_{ {\bf{q}}_{m}, {\bf{0}} }
\]
\begin{equation}
- \frac{ i\mbox{ }{\bf{q}}_{m}.{\bf{F}}([\rho];{\bf{q}}_{m}r) }
{ q_{m}^{2} }
\end{equation}
 In order to define $ X_{ {\bf{0}}r } $ in terms of fermi fields, we have
 to make use of the fact that this object does not commute with the total
 number of fermions. This means it cannot be expressed exclusively in terms
 of number-conserving fermi bilinears like currents and densities.
 This will mean that we merely invert the formula in Eq.(~\ref{FIELDOP})
 and solve for $ X_{ {\bf{0}}r } $ as,
\[
X_{ {\bf{0}}r } = -\sum_{ {\bf{k}}_{m} \neq {\bf{0}} }X_{ {\bf{k}}_{m}r }  +  
i\mbox{  }
\sum_{ {\bf{k}}_{m} }\mbox{  }ln[
 (\sqrt{ N_{r}^{0} } + 
\sum_{ {\bf{q_{n}}} } \delta{\mbox{ }}\psi({\bf{q_{n}}}r)T_{ -{\bf{q}}_{n} }
({\bf{k}}_{m}) )
(N_{r}^{0} + \sum_{ {\bf{q}}_{n} }\delta \rho({\bf{q}}_{n}r)T_{ {\bf{q}}_{n} }
({\bf{k}}_{m}))^{-\frac{1}{2}}exp(-i\mbox{ }\sum_{ {\bf{q}}_{n} }
\]
\begin{equation}
\times 
\phi([\rho];{\bf{q}}_{n}r)T_{ {\bf{q}}_{n} }({\bf{k}}_{m}))]
\mbox{    }\delta_{ {\bf{k}}_{m}, {\bf{0}} } 
\end{equation}

Define an operator which is defined to be the formal expansion
 that the formula itself suggests,
\begin{equation}
{\tilde{\psi}}({\bf{k}}_{n}r) =
exp(-i{\mbox{  }}
\sum_{ {\bf{q}}_{m}  }T_{ -{\bf{q}}_{m} }({\bf{k}}_{n})
X_{ {\bf{q}}_{m}r })
\mbox{ }exp(i{\mbox{  }}
\sum_{ {\bf{q}}_{m} }T_{ {\bf{q}}_{m} }({\bf{k}}_{n})
\phi([\rho];{\bf{q}}_{m}r)){\mbox{  }}
(N^{0}_{r} + \sum_{ {\bf{q}}_{m} }\delta\rho({\bf{q}}_{m}r)
T_{ {\bf{q}}_{m} }({\bf{k}}_{n}))^{\frac{1}{2}}
\delta_{ {\bf{k}}_{n}, {\bf{0}} }
\end{equation}
We would now like to write to write down a statement that
would require a proof.
This conjecture when proven will vindicate the DPVA.
\newline
{\bf{Conjecture}}
\newline
There exists a unique functional $ \Phi([\rho];{\bf{x}}r) $ and a unique odd
(for fermions, even for bosons)
 integer $ m $ such that the following recursion holds,
\[
\Phi([\{\rho({\bf{y_{1}}}\sigma_{1})
 - \delta({\bf{y_{1}}}-{\bf{x}}^{'})\delta_{\sigma_{1},\sigma^{'}} \} ]
;{\bf{x}}\sigma)
\]
\[
+ \Phi([\rho];{\bf{x^{'}}}\sigma^{'}) - \Phi([\rho];{\bf{x}}\sigma)
\]
\begin{equation}
-\Phi([\{\rho({\bf{y_{1}}}\sigma_{1})
 - \delta({\bf{y_{1}}}-{\bf{x}})\delta_{\sigma_{1},\sigma} \} ]
;{\bf{x^{'}}}\sigma^{'})
 = m\pi
\end{equation}
 and has the following additional effects.
The domain of definition of $ {\tilde{\psi}}({\bf{k}}_{n}r) $
( in which the series expansion converges ), is the same as that of
 $ \psi({\bf{k}}_{n}r) $ and it acts the same way too. In other words,
\begin{equation}
{\tilde{\psi}}({\bf{k}}_{n}r) = \psi({\bf{k}}_{n}r)
\end{equation}
 We know how the ingredients of $ {\tilde{\psi}}({\bf{k}}_{n}r) $ namely
 the current $ {\bf{j}}({\bf{k}}_{n}r) $ and the density
 $ \delta\rho({\bf{q}}_{n}r) $  act on typical elements of the Fock space,
 and we know how $  \psi({\bf{k}}_{n}r) $ acts on the Fock space,
 we just have to show that the complicated $ {\tilde{\psi}}({\bf{k}}_{n}r) $
 acts the same way as the simple $ \psi({\bf{k}}_{n}r) $. Moreover,
 this is true for a unique phase functional $ \Phi $.  Lastly, we would
 like to defend the above ``fourier gymnastics'' by pointing out
 that the real space formulation is not well-defined  due to
 the fact that the line-integral that appears in the formulas 
 is difficult to define, any attempt is equivalent to the above
 approach.  The other reason for attempting a rigorous formulation
 is the fact well-known to mathematicians that it is not possible
 to have a self-adjoint canonical conjugate of a positive definite
 self-adjoint operator. Since $ \rho $ is positive definite,
 the natural question that arises is whether $ \Pi $ self-adjoint ?
 We take the naive physicist's approach to this issue, namely
 we allow for sign changes in $ \rho $ and argue that these merely
 amount to translating the phase functional $ \Phi $ by constant amounts,
 thus not altering the general framework. Within this framework,
 $ \Pi $ is indeed self-adjoint and all is well. It is also worth remarking
 that the overall conjugate $ \Pi $ has two contributions, one from
 the position independent part $ X_{0\sigma} $, and the other is from 
 terms involving currents and densities. The latter contribution is 
 manifestly self-adjoint. The possible lack of self-adjointness of the overall
 conjugate stems from the canonical conjugate of the total number which
 cannot be expressed in terms of fermi bilinears.

\section{ACKNOWLEDGEMENTS}
 It is a pleasure to thank Prof. A. H. Castro-Neto and Prof. D. K. Campbell
 for providing important references and encouragement and the
 former for useful discussions as well, and
 Prof. A. J. Leggett for giving his valuable time and advice
 on matters related to the pursuit of this work and
 also for providing important references and also for
 useful discussions. Thanks are also due to
 Prof. Ilias E. Perakis for providing the authors with an important
 reference. Thanks are also due to
 Prof. P.W. Anderson for critically evaluating an early version
 of this article.  Last but not least to Dr. S. Chitanvis for correcting
 the authors' misreading of the Lieb-Mattis solution.
 This work was supported in part by ONR N00014-90-J-1267 and
 the Unversity of Illinois, Materials Research Laboratory under grant
 NSF/DMR-89-20539 and in part by the Dept. of Physics at
 University of Illinois at Urbana-Champaign. The authors may be contacted at
 the e-mail address setlur@mrlxpa.mrl.uiuc.edu.

\end{document}